\documentclass[aps,pre,onecolumn,amssymb,12pt]{revtex4-1}
\usepackage{color}
\usepackage{hyperref}
\hypersetup{
setpagesize=true,
 bookmarksnumbered=true,%
 bookmarksopen=true,%
 colorlinks=true,%
 linkcolor=blue,
 citecolor=red,
}
\usepackage{graphicx}
\usepackage{etoolbox}
\usepackage{bm}
\usepackage{amsmath}
\usepackage{lipsum}

\newcommand{\kt}{k_\mathrm{B}T}
\def\be{\begin{equation}}
\def\en{\end{equation}}
\def\p{\partial}

\def\ulq{\underline{q}}

\def\hlamb{\hat\lambda}
\def\ve{\varepsilon}

\def\nn{\nonumber}

\newcommand{\av}[1]{\langle{#1}\rangle}
\newcommand{\bav}[1]{\Big\langle{#1}\Big\rangle}

\begin{document}
\title{Fluctuating hydrodynamics of dilute electrolyte solutions: systematic perturbation calculation of effective transport coefficients governing large-scale dynamics}
\author{Ryuichi Okamoto}
\email{okamoto-ryuichi@okayama-u.ac.jp}
\affiliation{Research Institute for Interdisciplinary Science, Okayama University, Okayama 700-8530, Japan }
\date{\today}
\begin{abstract}
    We study the transport properties of dilute electrolyte solutions on the basis of the fluctuating hydrodynamic equation, which is a set of nonlinear Langevin equations for the ion densities and flow velocity. 
   The nonlinearity of the Langevin equations generally 
 leads to effective kinetic coefficients for the deterministic dynamics of the average ion densities and flow velocity; the effective coefficients generally differ from the counterparts in the Langevin equations and are frequency-dependent. Using the path-integral formalism involving auxiliary fields, we perform systematic perturbation calculations of the effective kinetic coefficients for ion diffusion, shear viscosity, and electrical conductivity, which  govern the dynamics on the large length scales. As novel contributions, we study the frequency dependence of the viscosity and conductivity in the one-loop approximation. Regarding the conductivity at finite frequencies, we derive the so-called electrophoretic part in addition to the relaxation part, where the latter has originally been obtained by Debye and Falkenhagen; it is predicted that the combination of these two parts gives rise to the frequency $\omega_{\rm max}$ proportional to the salt density, at which the real part of the conductivity exhibits a maximum. The zero-frequency limits of the conductivity and shear viscosity coincide with the classical limiting laws for dilute solutions, derived in different means by Debye, Falkenhagen, and Onsager. 
    As for the effective kinetic coefficients for slow ion diffusions in large length scales, our straightforward calculation yields the cross kinetic coefficient between cations and anions. Further, we discuss the possibility of extending the present study to more concentrated solutions.
\end{abstract}
\maketitle

\section{Introduction\label{sec:intro}}
The dynamic properties of electrolyte solutions, such as ionic diffusivity, fluidity, and electrical conductivity, are not only of fundamental interest in physics but are also important in electrochemistry\cite{Bockris2006} and nanofluidics\cite{Daiguji2010,Bocquet2010}. It is empirically known that some ion-specificities in the dynamic properties correlate with static, thermodynamic properties of chemical and biological systems, e.g., the Hofmeister series \cite{Kunz2004,Ninham2012}. In mixture systems, where the coupling between ion densities and solvent composition can be much greater than the thermal energy $\kt$, an addition of ions has more complex effects on both statics and dynamics, e.g., ion-induced (macro- or micro-) phase separations\cite{Sadakane2009,Okamoto2010,Onuki2011,Onuki2016} and colloidal aggregation\cite{Okamoto2011}.

The study of electrolyte solutions has a long history, but the understanding of ionic effects in solutions is far from complete, even in single-component solvent systems. In early years, several limiting (asymptotic) laws for dilute solutions were found, such as the Onsager limiting law of the electrical conductivity under DC electric field\cite{Onsager1927,RobinsonStokes} and the Falkenhagen-Fuoss-Onsager limiting law for the (effective) shear viscosity\cite{falkenhagenLXIIViscosityStrong1932,onsagerIrreversibleProcessesElectrolytes1932,OnsagerKim1957}. Later, calculations based on the mode-coupling theory (MCT) were carried out for solutions of higher concentrations, incorporating relaxation of the ionic atmosphere and hydrodynamic effects\cite{ChandraBagchi1999,ChandraBagchi2000_2,ChandraBagchi2000_3,Aburto2012,Aburto2013,Aburto2013_2}. 
Yamaguchi \textit{et al.} also proposed a generalized Langevin theory accounting for the contact ion pair\cite{Yamaguchi2007,Yamaguchi2009}. 


Another theoretical approach is the fluctuating hydrodynamics, where ion densities and the flow velocity are regarded as coarse-grained, continuous fields fluctuating owing to thermal noises\cite{Wada2005,Peraud2017,Donev2019}. 
Let $n_+({\bm r},t)$ and $n_-({\bm r},t)$ denote the coarse-grained number densities of the cations and anions, respectively, which depend on the position ${\bm r}$ and time $t$. These are assumed to obey the following Langevin equation including the fluid velocity ${\bm v}$ and the noise term $\eta_i$ ($i=+,-$):
\begin{align}
\frac{\p n_i}{\p t}=-{\bm v}\cdot \nabla n_i+\nabla\cdot \Big[ L_i \nabla \frac{\delta ({\cal F}/k_{\rm B}T)}{\delta n_i}\Big] +\eta_i, \label{diffusion}
\end{align}
where ${\cal F}[n_+,n_-]$ is the free energy functional for the ions and the noise correlation is given below by Eq.~(\ref{noise_corr}). Strictly speaking, the stochastic partial differential equations with infinite degrees of freedom like Eq.~(\ref{diffusion}) have no rigorous mathematical foundation unless one makes truncation/regularization/modification by introducing cutoff lengths, but we shall regard equations like Eq.~(\ref{diffusion}) as if they were Langevin equations of finite degrees of freedom as in many of the physics literature.
As we shall see later, the kinetic coefficient $L_i$ depends on $n_i$ and thus Eq.~(\ref{diffusion}) is a nonlinear Langevin equation with a multiplicative noise, of which exact treatment is difficult.The velocity field ${\bm v}$ is assumed to obey the incompressibility condition
\begin{align}
\nabla \cdot {\bm v}=0,
\end{align}
and the (generalized) Stokes equation with a noise term,
\begin{align}
    \frac{\p \rho{\bm v}}{\p t}=-\Big[ \sum_i n_i\nabla \frac{\delta {\cal F}}{\delta n_i} \Big]_\perp +\eta_0\nabla^2{\bm v}+{\bm f}, \label{deqv}
\end{align}
where $\rho$ is the mass density, $[\cdots]_\perp$ denotes the transverse mode, and $\eta_0$ is the shear viscosity of the pure solvent. The noise ${\bm f}$ satisfies \cite{LandauStat1,LandauStat2}
\begin{align}
\av{f_\alpha({\bm r},t)f_\beta({\bm r}',t')}=-2\eta_0 T (\delta_{\alpha\beta}\nabla^2-\nabla_\alpha\nabla_\beta)\delta_{{\bm r}-{\bm r}'}\delta_{t-t'}, \label{f_corr}
\end{align}
where $\delta_{{\bm r}-{\bm r}'}=\delta({\bm r}-{\bm r}')$ and $\delta_{t-t'}=\delta(t-t')$ are the delta functions.
Hereafter the greek subscript (superscript) letters denote the spacial coordinate, i.e., $x$, $y$, and $z$, while the roman subscript (superscript) letters denote the indices of the scalar fields, i.e., $+$, $-$ or $1$, $2$.

Linearizing the fluctuating hydrodynamic equations (FHE) with respect to the deviations of the fields from their respective average values and neglecting the multiplicative part of the noises, Wada studied the fluctuation contribution to the shear viscosity and derived the Falkenhagen-Fuoss-Onsager limiting law for symmetric salts that the cations and anions have the same diffusivity\cite{Wada2005}. More recently, P\'eraud \textit{et al.} used the linearized FHE to derive the Onsager limiting law for conductivity for symmetric salts. Soon later, the same authors generalized their calculation to asymmetric salts and symmetric ternary salts\cite{Donev2019}. In a similar manner, D\'emery and Dean\cite{Demery2016} also studied the nonlinear response to the external electric field (Wien effect) and derived the field-dependent conductivity originally derived by Onsager and Kim\cite{OnsagerKim1957_2}, while the fluid velocity was neglected.
The FHE can also be used for numerical simulations of equilibrium and nonequilibrium situations of a large system with relatively low computational cost\cite{Peraud2016}. One may also use the FHE for more concentrated solutions, taking into account the inter-ionic short-range interactions. Such short-range interactions are responsible for the ion-specificity of thermodynamics of electrolyte solutions, and are explained by means of the interplay of the steric effect (among the solvent and ions) and the electrostriction due to ions \cite{Okamoto2020,Okamoto2021}. Recently, Avni \textit{et al.} have incorporated the hard-sphere-like repulsion between ions in the FHE to study the ion-specific DC conductivity in concentrated electrolyte solutions, and their result fits the experimental data well up to the salt concentration of 3 molars \cite{Avni2022}.

Meanwhile, deterministic equations which are the same as Eqs.~(\ref{diffusion}) and (\ref{deqv}) but \textit{without the noise terms} have extensively been used in many research fields such as nanofluidics\cite{Daiguji2010} and colloidal science\cite{Russel1991colloidal} to study the time evolution of the average ion densities and flow velocity. However, it is well-known that the nonlinearity of the Langevin equations does not generally allow one to simply discard the noise terms to have the deterministic equations for the averaged variables because the averaged products of fluctuating variables are not equal to the products of averaged variables. In fact, for the averaged variables, the fluctuation effect leads to the \textit{effective} kinetic coefficients that differ from the counterparts in the Langevin equations, e.g., $L_i$ in Eq.~(\ref{diffusion}) and $\eta_0$ in Eq.~(\ref{deqv}). 
Reflecting the collective motions of ions and solvent molecules on the meso-scales, the effective coefficients are generally frequency- and wavenumber-dependent. For non-ionic solutions, Bedeaux and Mazur have investigated in detail such effective diffusion coefficient by incorporating the fluid velocity, using nonlinear fluctuating hydrodynamics\cite{Bedeaux1974,Mazur1974}.
As shall be more apparent in the following, the above-mentioned shear viscosity of electrolyte solutions is one of those effective coefficients in the limit of zero frequency and wavenumber. Note that the magnitude of fluctuation effects in the viscosity and the conductivity is large enough to be detected in experiments\cite{RobinsonStokes, JenkinsMarcus1995}. Therefore, theoretical investigations of fluctuation effects on the transport coefficients are not only of fundamental interest, but also important for applications in many research fields.

In this paper, we derive the effective kinetic coefficients for ion diffusion [the counterpart of $L_i$ in Eq.~(\ref{diffusion})], and the frequency-dependent shear viscosity and electrical conductivity, which are missing in the literature (To be precise, as for the conductivity, the frequency dependence of the so-called relaxation part has already been studied by Debye and Falkenhagen\cite{DebyeFalken}, but none has derived the electrophoretic part for finite frequencies.). The frequency-dependent coefficients reduce to the classical limiting laws\cite{Onsager1927,falkenhagenLXIIViscosityStrong1932,onsagerIrreversibleProcessesElectrolytes1932} at zero frequency. We shall see that the combination of relaxation and electrophoretic parts gives rise to a maximum in the real part of the conductivity at the frequency $\omega_{\rm max}$ proportional to the salt density.
We use the path-integral formalism developed by Martin, Siggia, Rose, Janssen, and de Dominicis (MSRJD)\cite{MSR1973,Janssen1976,DeDominicis1976}, which naturally leads to the effective kinetic coefficients and also enables us to perform systematic perturbation calculations; although in this paper we only perform the lowest order (one-loop) calculation, one can in principle study higher-order effects. We shall also discuss without performing explicit calculations how a two-loop term and inter-ionic short-range interactions\cite{Okamoto2020}, both of which would be relevant for more concentrated solutions, contribute to the effective shear viscosity. 

The organization of the paper is as follows. In Sec.~\ref{sec:theory} we introduce the so-called Janssen-De Dominicis action corresponding to our (nonlinear) FHE. We then discuss how the effective transport coefficients are expressed in terms of correlation functions and vertex functions. 
In Sec.~\ref{sec:result} transport properties are discussed in the one-loop approximation. We first derive the effective kinetic coefficients associated with the slow dynamics of long-wavelength ion densities by calculating the corresponding vertex functions; they naturally lead to the renormalized diffusion coefficients\cite{Peraud2017,Donev2019} into which cutoff-sensitive fluctuation corrections are absorbed. We then investigate the effective shear viscosity and the frequency dependence of the electrical conductivity. In Sec.~\ref{sec:discussion} we discuss the higher-order effects, avoiding explicit calculations. Section \ref{sec:summary} is devoted for summary.

\section{Theoretical background\label{sec:theory}}
For the sake of lighter notation, we hereafter take the units such that the thermal energy $k_{\rm B}T$ is denoted by $T$.
We assume that the noise $\eta_i$ in Eq.~(\ref{diffusion}) is a Gaussian white noise satisfying
\begin{align}
\av{\eta_i({\bm r},t)\eta_j({\bm r}',t')}=2\delta_{ij}\nabla\nabla' L_i(n_i({\bm r},t))\delta_{{\bm r}-{\bm r}'}\delta_{t-t'}. \label{noise_corr}
\end{align}
In the above, notice that the noise $\eta_i$ is multiplicative because of the dependence of $L_i$ on $n_i$.
The free energy functional ${\cal F}$ (within the local approximation) is written in the form:
\begin{align}
{\cal F}=\int d{\bm r} \Big[ f_{\rm loc}(n_+, n_-)+\frac{\ve (\nabla \Psi)^2}{8\pi} \Big] \label{FreeEnergy}
\end{align}
where $f_{\rm loc}$ is the local free energy density, and the second term in the integrand is the electrostatic energy. 
The electrostatic potential $\Psi$ is a functional of the ion densities $n_i$ determined by the Poisson equation
\begin{align}
- \ve \nabla^2 \Psi=4\pi e\sum_{i=+,-}Z_in_i, \label{Poisson}
\end{align}
where $Z_i$ is the valence number of the species $i$, $e$ the elementary charge, and $\ve$ the dielectric permittivity of the solvent.
It is natural that in Eq.~(\ref{diffusion}) the combination of $L_i$ and the entropic term of ${\cal F}$ yields the diffusion term $D_i\nabla^2 n_i$ with the (bare) diffusion coefficient $D_i$ of species $i$ [See Eq.~(\ref{dilute})]. Hence the following is usually assumed:
\begin{align}
    L_i=D_i n_i({\bm r},t). \label{kinetic_D}
\end{align}

Generally, the nonlinear terms in Eq.~(\ref{diffusion}) and the multiplicative nature of the noise make the mathematical treatment difficult, and hence numerical and/or approximation is necessary.
In the previous theories based on the FHE \cite{Wada2005,Peraud2017,Donev2019}, where the limiting laws in dilute solutions are discussed, the authors made (at least) two assumptions. First, the local free energy density in Eq.~(\ref{FreeEnergy}) is given by only the entropic contribution for dilute solutions:
\begin{align}
f_{\rm loc}(n_+, n_-)\approx T\sum_{i=\pm} n_i[  \ln(\lambda_i^3 n_i)-1],\label{dilute}
\end{align}
where $\lambda_i$ is the thermal de Broglie length of species $i$.
Second, the fluctuations are small so that the Langevin equation (\ref{diffusion}) can be linearized with respect to ${\bm v}-\av{{\bm v}}$ and the density deviation $\delta n_i =n_i -\av{n_i}$, and accordingly $n_i$ in Eq.~(\ref{kinetic_D}) is replaced by its average value $\av{n_i}$. The linearization is coherent with the first assumption because in dilute solutions the noise strength for ions is small as can be seen from Eqs.~(\ref{noise_corr}) and (\ref{kinetic_D}). Beyond the limiting regime, i.e., for slightly more concentrated solutions, one cannot neglect the nonlinear terms in Eq.~(\ref{diffusion}) since the noise becomes stronger. Furthermore, the interaction term $(1/2)\sum_{i,j=\pm}U_{ij}^{\rm eff}n_in_j$ should also be included in $f_{\rm loc}$. Here, the interaction coefficient $U_{ij}^{\rm eff}$, which is accounting for both the inter-ionic direct interaction and the solvent-mediated interaction, is highly ion-specific\cite{Okamoto2018,Okamoto2020,Okamoto2021}.

In the present study of dilute solutions, we remain using the free energy density in Eq.~(\ref{dilute}). Then the functional derivative in Eq.~(\ref{diffusion}) is calculated as
\begin{align}
\frac{\delta ({\cal F}/T)}{\delta n_i}=\ln(\lambda_i^3 n_i)+Z_iU. \label{dF}
\end{align}
where we have defined the normalized electrostatic potential $U\equiv e\Psi/T$. 
For simplicity, we assume that the salt is monovalent, i.e.,
\begin{align}
    Z_+=-Z_-=-1.
\end{align}
Thus the average ion densities are given by the salt density $\bar n$,
\begin{align}
    \av{n_+}=\av{n_-}=\bar n.
\end{align}


\subsection{Equations for transformed density variables}
It is convenient to introduce the ion density deviation $\phi_1$ and the charge density $\phi_2$ as
\begin{align}
\phi_1\equiv n_++n_--\bar n,\quad \phi_2\equiv n_+-n_-. \label{def:phi}
\end{align}
The Poisson equation (\ref{Poisson}) is then rewritten as
\begin{align}
    -\nabla^2 U=4\pi \ell_{\rm B}\phi_2,\quad  \label{Poisson2}
\end{align}
where $\ell_{\rm B}=e^2/T\ve$ is the Bjerrum length.
The change rate $\dot\phi_i=\p\phi/\p t$ is given by the sum of the deterministic part $g_i$ and the noise $\xi_i$,
\begin{align}
    \dot\phi_i=g_i+\xi_i. \label{dotphi}
\end{align}
Using Eqs.~(\ref{kinetic_D}), (\ref{dF}), and (\ref{def:phi}), we obtain
\begin{align}
    g_1=&-{\bm v}\cdot \nabla \phi_1+D[\nabla^2\phi_1+\nabla\cdot \phi_2\nabla U] +\gamma D[\nabla^2\phi_2+\nabla\cdot (\phi_1+2\bar n)\nabla U] \label{deq1}\\
g_2=&-{\bm v}\cdot \nabla \phi_2+D[\nabla^2\phi_2+\nabla\cdot (\phi_1+2\bar n)\nabla U] +\gamma D[\nabla^2\phi_1+\nabla\cdot \phi_2\nabla U] \label{deq2} \\
& \hspace{-6.8mm}\xi_1= \eta_++\eta_-,\quad \xi_2=\eta_+-\eta_-.
\end{align}
where we have defined the mean (bare) diffusion coefficient $D$ and (bare) asymmetry factor $\gamma$, 
\begin{align}
&D\equiv (D_++D_-)/2,\quad \gamma\equiv (D_+-D_-)/2D
\end{align}
Equations (\ref{noise_corr}) and (\ref{kinetic_D}) yield the noise correlation:
\begin{align}
    &\av{\xi_1(\underline r)\xi_2(\underline r')}=2D\nabla'\cdot\nabla  [\phi_2(\underline r)+\gamma \phi_1(\underline r)+2\gamma \bar n]\delta_{\underline r-\underline r'} \label{xi_corr1}\\
    &\av{\xi_i(\underline r)\xi_i(\underline r')}=2D\nabla'\cdot\nabla [\phi_1(\underline r)+\gamma\phi_2(\underline r)+2\bar n] \delta_{\underline r-\underline r'},  \label{xi_corr2}
    \end{align}
where we have used the abbreviation
\begin{align}
    \underline r\equiv({\bm r},t). \nn
\end{align}

In Eq.~(\ref{deqv}), we assume the mass density $\rho$ is constant in space and time for simplicity,
\begin{align}
    \rho=\mathrm{const.}
\end{align}
Using Eq.~(\ref{dF}), we can rewrite the first term of the right hand side of Eq.~(\ref{deqv}) as
\begin{align}
\Big[ \sum_i n_i\nabla \frac{\delta {\cal F}}{\delta n_i} \Big]_\perp=T [ \nabla \phi_1+\phi_2 \nabla U ]_\perp=T[\phi_2 \nabla U ]_\perp.
\end{align}
Equations (\ref{deqv}), (\ref{Poisson2}), and (\ref{dotphi}) with the noise correlations Eqs.~(\ref{f_corr}), (\ref{xi_corr1}), and (\ref{xi_corr2}) are our full FHE for the dynamics of $\phi_1$, $\phi_2$, and ${\bm v}$.

As has been pointed out below Eq.~(\ref{noise_corr}), Eq.~(\ref{diffusion}) [or (\ref{dotphi})] is a Langevin equation with a multiplicative noise. Hence we must specify the interpretation as a stochastic differential equation, i.e., It\^o, Stratonovich, or other\cite{vanKampen1992}. To the best of the author's knowledge, there is no general way to determine which interpretation to adopt for a given physical Langevin equation. However, Dean\cite{Dean1996} has formally derived a Langevin equation of the form of Eq.~(\ref{diffusion}) without the streaming term $-{\bm v}\cdot \nabla n_i$, applying the It\^o calculus to a system of Brownian particles interacting via a pairwise potential. In the present study we thus choose It\^o prescription, which also simplifies the subsequent analysis. One must also note that in general we need to add a deterministic (drift) term to the right hand side of Eq.~(\ref{diffusion}) (and hence to Eq.~(\ref{dotphi})), in order that the equilibrium probability distribution of the density fluctuations obeys the Boltzmann weight $\propto \exp(-{\cal F}/T)$; this additional term depends on the adopted interpretation of the Langevin equation \cite{OnukiBook,Lau2007}. In our case, we can show that the additional term vanishes for It\^o interpretation under the choice of $L_i$ in Eq.~(\ref{kinetic_D}) (see Appendix \ref{sec:drift}).

\subsection{Martin-Siggia-Rose-Janssen-De Dominicis formalism \label{sec:MSRJD}}
To study our FHE, which is nonlinear and with multiplicative noises, we use the path-integral formalism developed by Martin, Siggia, Rose, Janssen, and de Dominicis (MSRJD)\cite{MSR1973,Janssen1976,DeDominicis1976}, which has extensively been used to study the critical dynamics\cite{Tauber}. In this formalism, auxiliary fields $\tilde\phi_1$, $\tilde\phi_2$, and $\tilde{\bm v}$ conjugate to $\phi_1$, $\phi_2$, and ${\bm v}$, respectively, are introduced. The statistical average of a functional $A[\phi_1, \phi_2, {\bm v}]$ is expressed as the path-integral,
\begin{align}
    \av{A}={\cal C}^{-1}\int {\cal D}{\bm v} {\cal D}\tilde{\bm v} \prod_{i=1,2}[{\cal D}\phi_i{\cal D}\tilde\phi_i]\, e^{-S}A, \label{MSRJD}
\end{align}
where ${\cal C}\equiv \int {\cal D}{\bm v} {\cal D}\tilde{\bm v} \prod_i[{\cal D}\phi_i{\cal D}\tilde\phi_i]\, e^{-S}$ is the normalization constant, and the integrations with respect to the auxiliary fields are performed along the imaginary axis. While it is known that expressions like Eq.~(\ref{MSRJD}) are mathematically ill-defined for non-Gaussian actions, we use it for non-Gaussian action $S$ as in the most of the physics literature. The action $S$ in Eq.~(\ref{MSRJD}), which is often called Janssen-De Dominicis action, is given by
\begin{align}
    S=\int_{\underline r} \Big[ &\sum_i\tilde\phi_i(\dot\phi_i-g_i)  +\tilde{\bm v}\cdot \Big(\dot{\bm v}-\nu_0\nabla^2{\bm v}+\frac{T}{\rho}[\phi_2\nabla U]_\perp\Big) \nn\\
    &+2D\bar n \Big( \tilde\phi_1\nabla^2\tilde\phi_1 +\tilde\phi_2\nabla^2\tilde\phi_2+2\gamma\tilde\phi_1\nabla^2\tilde\phi_2\Big) \nn\\
    &-D\big\{ |\nabla \tilde\phi_1|^2(\phi_1+\gamma\phi_2)+|\nabla \tilde\phi_2|^2(\phi_1+\gamma\phi_2) +2(\nabla\tilde\phi_1)\cdot (\nabla\tilde\phi_2)(\phi_2+\gamma\phi_1)\big\} \nn\\
    &+T(\eta_0/\rho^2)\sum_{\alpha\beta}\tilde v_\alpha (\delta_{\alpha\beta}\nabla^2-\nabla_\alpha\nabla_\beta)\tilde v_\beta \Big],\label{action}
\end{align}
where we have introduced the abbreviation $\int_{\underline r}=\int d{\bm r}\int dt$, and the kinetic viscosity $\nu_0=\eta_0/\rho$ of the solvent.
The auxiliary fields are not stochastic variables, and neither is thus the auxiliary-fields-dependent functional $B[\phi_1, \phi_2, {\bm v},\tilde\phi_1,\tilde\phi_1,\tilde{\bm v}]$. Hence, if we replace $A[\phi_1, \phi_2, {\bm v}]$ by $B[\phi_1, \phi_2, {\bm v},\tilde\phi_1,\tilde\phi_1,\tilde{\bm v}]$ in the right hand side of Eq.~(\ref{MSRJD}), it cannot be interpreted as the average of $B$. However, for the sake of simplicity, we use the same notation $\av{\cdots}$ as if it were a statistical average of a fluctuating physical quantity,
\begin{align}
    \av{B}={\cal C}^{-1}\int {\cal D}{\bm v} {\cal D}\tilde{\bm v} \prod_{i=1,2}[{\cal D}\phi_i{\cal D}\tilde\phi_i]\, e^{-S}B,
\end{align}
and shall be called the average of $B$.

We divide the action $S$ into four parts
\begin{align}
S=S_0+S_{\phi\phi}+S_{\rm m}+S_{\phi v}, \label{action_sum}
\end{align}
where $S_0$ is the bilinear part, $S_{\phi\phi}$ is the non-linear coupling of $\phi_i$'s, $S_{\rm m}$ is stems from the multiplicative part of the noises, and $S_{\phi v}$ is from the reversible mode-coupling between $\phi_i$ and ${\bm v}$. In the following we shall write the action in terms of the Fourier components of the fields; the space-time Fourier transform of any field $g(\underline r)$ is defined as
\begin{align}
    g(\underline{q})\equiv \int_{\underline r} g(\underline r)e^{-i\omega t-i{\bm q}\cdot {\bm r}},  \quad \ulq\equiv ({\bm q},\omega) . \label{Fourier1}
\end{align}
In terms of the Fourier components, the Poisson equation in Eq.~(\ref{Poisson2}) is written as
\begin{align}
 U=4\pi\ell_{\rm B}\phi_2/q^2.\label{Poisson3}
\end{align}

\subsubsection{Bilinear action and propagators}
\begin{figure}[htp]
\includegraphics[width=0.7\columnwidth]{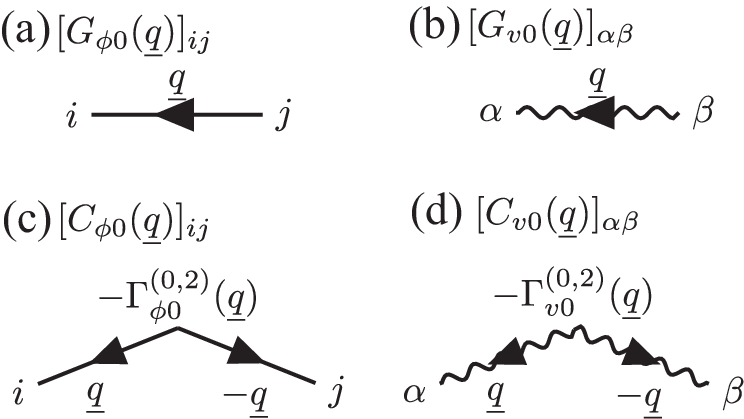}
\caption{Graph representations of the (a),(b) response propagators and (c),(d) correlation propagators. \label{fig:propagators}}
\end{figure}

We first discuss the bilinear part $S_0$, which yields the response and correlation propagators.
We define 
\begin{align}
    &\Gamma_{\phi 0}^{(1,1)}(\ulq)\equiv \begin{pmatrix} 
        i\omega+Dq^2 & \gamma D(q^2+\kappa^2) \\
        \gamma D q^2 & i\omega+D(q^2+\kappa^2)
        \end{pmatrix} \label{Gamma_p0_1}\\
    &\Gamma_{\phi 0}^{(0,2)}(\ulq) \equiv -4\bar n Dq^2 \begin{pmatrix}
        1 & \gamma \\
        \gamma & 1
        \end{pmatrix}. \label{Gamma_p0_2}
\end{align}
and 
\begin{align}
    &\bar\Gamma_{v 0}^{(1,1)}(\ulq)\equiv i\omega+\nu_0 q^2, \quad \Gamma_{v 0}^{(1,1)}(\ulq)\equiv \bar\Gamma_{v 0}^{(1,1)}(\ulq){\cal P}^\perp ({\bm q}) \\
&\bar\Gamma_{v 0}^{(0,2)}(\ulq)\equiv -2T(\eta_0/\rho^2)q^2,\quad \Gamma_{v 0}^{(0,2)}(\ulq)\equiv \bar\Gamma_{v 0}^{(0,2)}(\ulq){\cal P}^\perp ({\bm q}) \label{Gamma_v0_1}
\end{align}
where $[{\cal P}^\perp({\bm q})]_{\alpha\beta}\equiv {\cal P}_{\alpha\beta}^\perp({\bm q})\equiv \delta_{\alpha\beta}-q_\alpha q_\beta/q^2$ is the projection operator onto the plane perpendicular to ${\bm q}$. In terms of the matrices in Eqs.~(\ref{Gamma_p0_1})--(\ref{Gamma_p0_2}), we define $4\times 4$ and $6\times 6$ matrices,
\begin{align}
    &\Gamma_{\phi 0}(\ulq)\equiv \left( \begin{array}{c|c}
        0 & \Gamma_{\phi 0}^{(1,1)}(\ulq)^\dagger \\ \hline
        \Gamma_{\phi 0}^{(1,1)}(\ulq) & \Gamma_{\phi 0}^{(0,2)}(\ulq) 
        \end{array}\right)\\
    &\Gamma_{v 0}(\ulq)\equiv \left( \begin{array}{c|c}
        0 & \Gamma_{v 0}^{(1,1)}(\ulq)^\dagger \\ \hline
        \Gamma_{v 0}^{(1,1)}(\ulq) & \Gamma_{v 0}^{(0,2)}(\ulq) 
        \end{array}\right),
\end{align}
and the vectors,
\begin{align}
    &\Phi(\ulq)\equiv (\phi_1(\ulq), \phi_2(\ulq), \tilde\phi_1(\ulq),\tilde\phi_2(\ulq))^t \\
    & V(\ulq) \equiv (v_x(\ulq),v_y(\ulq),v_z(\ulq), \tilde v_x(\ulq),\tilde v_y(\ulq),\tilde v_z(\ulq))^t,
 \end{align}
 where $\dagger$ and $t$ denote the Hermite conjugate and the transpose, respectively.
The bilinear part $S_0$ is then written as
\begin{align}
    S_0=\frac{1}{2}\int_{\underline{q}} \big[\Phi(-\ulq)\Gamma_{\phi 0}(\ulq)\Phi(\ulq)+V(-\ulq)\Gamma_{v 0}(\ulq)V(\ulq)\big], \label{S0}
\end{align}
where we have introduced the abbreviation $\int_{\ulq}=(2\pi)^{-4}\int d\omega\int d{\bm q}$.
Equation (\ref{S0}) yields the response propagators
\begin{align}
    & G_{\phi 0}(\ulq)=[\Gamma_{\phi 0}^{(1,1)}(\ulq)]^{-1} \label{G_p0}\\
    & G_{v 0}(\ulq)={\cal P}^\perp({\bm q})/\bar\Gamma_{v0}^{(1,1)}(\ulq),
\end{align}
and the correlation propagators
\begin{align}
    &C_{\phi 0}(\ulq)=-G_{\phi 0}(\ulq)\Gamma_{\phi 0}^{(0,2)}(\ulq)G_{\phi 0}(\ulq)^\dagger \label{C_p0}\\
    &C_{v0}(\ulq)=-G_{v0}(\ulq)\Gamma_{v0}^{(0,2)}(\ulq)G_{v0}(\ulq)^\dagger. \label{C_v0}
\end{align}
Obviously, the Gaussian averages $\av{\cdots}_0={\cal C}_0^{-1} \int {\cal D}{\bm v} {\cal D}\tilde{\bm v} \prod_{i=1,2}[{\cal D}\phi_i{\cal D}\tilde\phi_i]e^{-S_0}(\cdots)$ of the pairs of one physical and one auxiliary fields yield the response propagators where ${\cal C}_0$ is the normalization constant. That is, $\av{\phi_i(\ulq_1)\tilde\phi_j(\ulq_2)}_0=(2\pi)^4\delta_{\ulq_1+\ulq_2}[G_{\phi 0}(\ulq_1)]_{ij}$ and $\av{v_\alpha(\ulq_1)\tilde v_\beta(\ulq_2)}_0=(2\pi)^4\delta_{\ulq_1+\ulq_2}[G_{v 0}(\ulq_1)]_{\alpha\beta}$. Similarly, the averages of the two physical fields yield the correlation propagator, $\av{\phi_i(\ulq_1)\phi_j(\ulq_2)}_0=(2\pi)^4\delta_{\ulq_1+\ulq_2}[C_{\phi 0}(\ulq_1)]_{ij}$ and $\av{v_\alpha(\ulq_1) v_\beta(\ulq_2)}_0=(2\pi)^4\delta_{\ulq_1+\ulq_2}[C_{v 0}(\ulq_1)]_{\alpha\beta}$.
Figure \ref{fig:propagators} presents the graph-representations of the propagators. The response propagators $G_{\phi 0}$ and $G_{v0}$ are represented by a directed solid line and by a directed wavy line, respectively (Fig.~\ref{fig:propagators}a,b). The correlation propagators in Eqs.~(\ref{C_p0}) and (\ref{C_v0}) are respectively represented by two response propagators outgoing from the vertices of $-\Gamma_{\phi 0}^{0,2}$ and $-\Gamma_{v 0}^{0,2}$ (Fig.~\ref{fig:propagators}c,d).

\subsubsection{Nonlinear parts}
\begin{figure}[htp]
    \includegraphics[width=0.7\columnwidth]{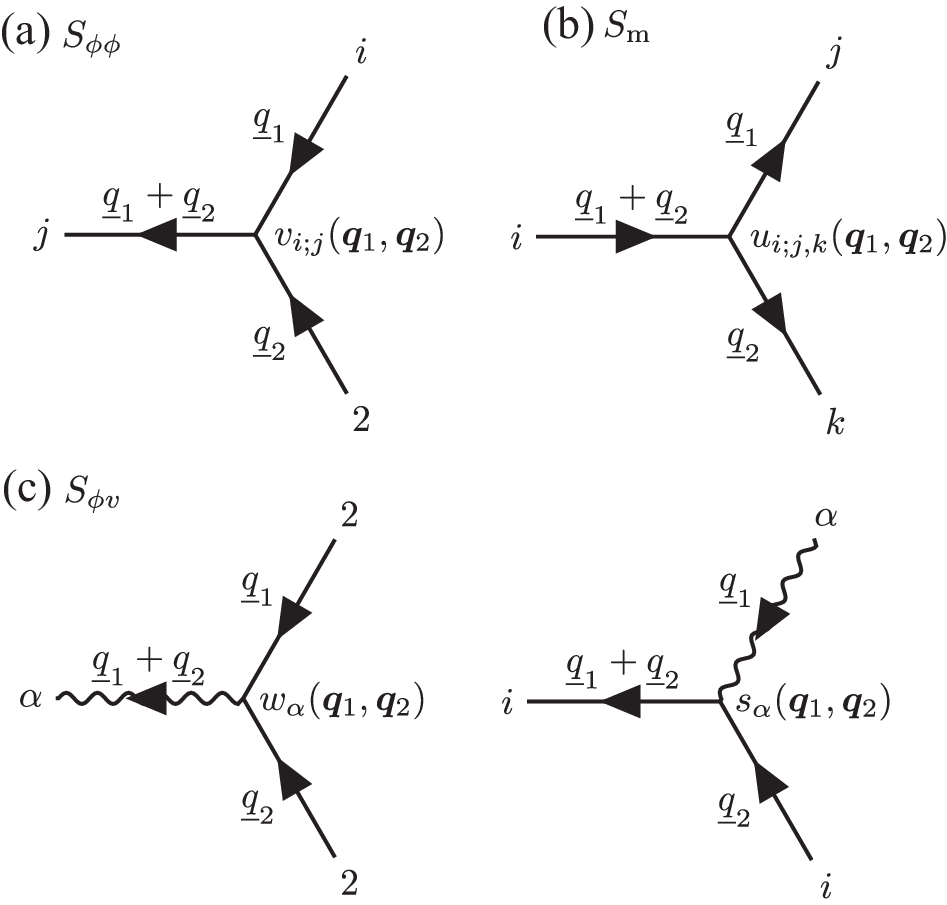}
    \caption{Graph representations of the interaction terms in (a)$S_{\phi\phi}$, (b)$S_{\rm m}$, and (c)$S_{\phi v}$. \label{fig:vertices}}
    \end{figure}

Now we discuss the nonlinear parts in Eq.~(\ref{action}).
First, the nonlinear part $S_{\phi\phi}$ arises from nonlinear coupling of $\phi_i$'s in the Langevin equation. From Eqs.~(\ref{deq1}), (\ref{deq2}), (\ref{action}) and (\ref{Poisson3}), we obtain
\begin{align}
    S_{\phi\phi}=\int_{\ulq_1,\ulq_2}\sum_{i,j}\frac{v_{i;j}({\bm q}_1,{\bm q}_2)}{1+\delta_{2i}}\phi_i(\ulq_1)\phi_2(\ulq_2)\tilde\phi_j(-\ulq_1-\ulq_2),
\end{align}
where the vertex factors $v_{i;j}$ are given by
\begin{align}
    &v_{1;1}({\bm q}_1,{\bm q}_2)=4\pi\ell_{\rm B}\gamma Dq_2^{-2}{\bm q}_2\cdot({\bm q}_1+{\bm q}_2) \label{v_ij1}\\
    &v_{1;2}({\bm q}_1,{\bm q}_2)=4\pi\ell_{\rm B}Dq_2^{-2}{\bm q}_2\cdot({\bm q}_1+{\bm q}_2)\\
    &v_{2;1}({\bm q}_1,{\bm q}_2)=4\pi\ell_{\rm B}D(q_1^{-2}{\bm q}_1+q_2^{-2}{\bm q}_2)\cdot({\bm q}_1+{\bm q}_2) \\
    &v_{2;2}({\bm q}_1,{\bm q}_2)=4\pi\ell_{\rm B}\gamma D(q_1^{-2}{\bm q}_1+q_2^{-2}{\bm q}_2)\cdot({\bm q}_1+{\bm q}_2) \label{v_ij4}.
 \end{align}
Similarly, we have the contribution from the multiplicative noise,
\begin{align}
    S_{\rm m}=\int_{\ulq_1,\ulq_2}\sum_{i,j,k}\frac{u_{i;jk}({\bm q}_1,{\bm q}_2)}{2}\phi_i(-\ulq_1-\ulq_2)\tilde\phi_j(\ulq_1)\tilde\phi_k(\ulq_1)
\end{align}
with the vertex factors
\begin{align}
    &u_{1;11}({\bm q}_1,{\bm q}_2)=u_{1;22}({\bm q}_1,{\bm q}_2) =u_{2;12}({\bm q}_1,{\bm q}_2)=u_{2;21}({\bm q}_1,{\bm q}_2)=2D{\bm q}_1\cdot{\bm q}_2  \label{vertex_u1}\\
    &u_{2;11}({\bm q}_1,{\bm q}_2)=u_{2;22}({\bm q}_1,{\bm q}_2)=u_{1;12}({\bm q}_1,{\bm q}_2)=u_{1;21}({\bm q}_1,{\bm q}_2)=2\gamma D{\bm q}_1\cdot{\bm q}_2.\label{vertex_u2}
\end{align}
Finally, the part $S_{\phi v}$ due to the reversible mode-coupling terms is obtained as
\begin{align}
    S_{\phi v}=\int_{\ulq_1,\ulq_2}\sum_\alpha\Big[& \frac{w_\alpha({\bm q}_1,{\bm q}_2)}{2}\phi_2(\ulq_1)\phi_2(\ulq_2)\tilde v_\alpha(-\ulq_1-\ulq_2) \nn\\
    &+\sum_i s_\alpha({\bm q}_1,{\bm q}_2) v_{\alpha}(\ulq_1)\phi_i(\ulq_2)\tilde\phi_i(-\ulq_1-\ulq_2)\Big]
\end{align}
with 
\begin{align}
    &w_\alpha({\bm q}_1,{\bm q}_2)=\frac{i4\pi\ell_{\rm B}T}{\rho}[ {\cal P}^\perp({\bm q}_1+{\bm q}_2)(q_1^{-2}{\bm q}_1+q_2^{-2}{\bm q}_2)]_\alpha \label{vertex_w}\\
    &s_\alpha({\bm q}_1,{\bm q}_2)=i[{\cal P}^\perp({\bm q}_1){\bm q}_2]_\alpha. \label{vertex_s}
\end{align}
The terms in $S_{\phi\phi}$, $S_{\phi v}$, $S_{\rm m}$, and $S_{\phi v}^\alpha$ are represented by graphs (vertices) shown in Fig.~\ref{fig:vertices}. The incoming and outgoing lines in the graphs represent the physical fields and auxiliary fields, respectively.

\subsection{Vertex functions and effective kinetic coefficients \label{sec:vertex}}
We start with rather trivial relations for the linearized equations. Linearizing the Langevin equation Eq.~(\ref{dotphi}) with respect to $\phi_i$ and ${\bm v}$, one readily obtains
\begin{align}
    \Gamma_{\phi 0}^{(1,1)}(\ulq) \begin{pmatrix} 
        \phi_1(\ulq)  \\
        \phi_2(\ulq) 
        \end{pmatrix} =\begin{pmatrix} 
            \xi_1(\ulq)  \\
            \xi_2(\ulq) 
            \end{pmatrix} \label{Lin_Lan}
\end{align}
Accordingly, the noise correlation without the multiplicative part is give by
\begin{align}
    \av{\xi_i(\ulq_1)\xi_j(\ulq_2)}_0 =-(2\pi)^4[\Gamma_{\phi 0}^{(0,2)}(\ulq_1)]_{ij}\delta_{\ulq_1+\ulq_2} \label{Lin_noise}
\end{align}
Equation (\ref{Lin_Lan}) indicates that (at the linear level) the response to the external field is governed by $\Gamma_{\phi 0}^{(1,1)}$, while Eq.~(\ref{Lin_noise}) indicates the kinetic coefficients associated with $\phi_i$ is given by
\begin{align}
    {\cal L}_{ij}=-\frac{1}{2}\frac{\p [\Gamma_{\phi 0}^{(0,2)}(\ulq)]_{ij}}{\p (q^2)} \Big|_{({\bm q},\omega)=(0,0)}, \label{kinetic0}
\end{align}
which yields
\begin{align}
    {\cal L}_{11}={\cal L}_{22}=2\bar nD,\quad {\cal L}_{12}={\cal L}_{21}=2\bar n \gamma D. \label{Lij_0loop}
\end{align}
Of course, the kinetic coefficients associated with $n_+$ and $n_-$ are recovered from the trivial relations $L_{++}=({\cal L}_{11}+{\cal L}_{22}+2{\cal L}_{12})/4$, $L_{--}=({\cal L}_{11}+{\cal L}_{22}-2{\cal L}_{12})/4$ and $L_{+-}=({\cal L}_{11}-{\cal L}_{22})/4$ as
\begin{align}
    L_{++}=D_+\bar n,\quad L_{--}=D_- \bar n,\quad L_{+-}=L_{-+}=0.
\end{align}
Here, $L_{ii}$ are indeed the same as $L_i$ in Eq.~(\ref{kinetic_D}) with $n_i$ being replaced by $\bar n$. Similarly we can express the solvent viscosity in terms of $\bar\Gamma_{v0}^{(0,2)}$ as
\begin{align}
    \eta_0=-(\rho^2/2T) \frac{\p \bar\Gamma_{v 0}^{(0,2)}(\ulq)}{\p (q^2)} \Big|_{({\bm q},\omega)=(0,0)} \label{eta0_vertex}
\end{align}

\begin{figure}[htp]
    \includegraphics[width=0.7\columnwidth]{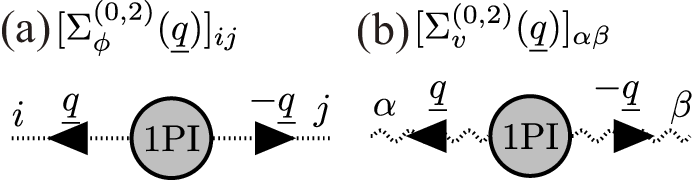}
    \caption{Graph representations of the self-energies (a)$\Sigma_{\phi}^{(0,2)}(\ulq)$ and (b)$\Sigma_{v}^{(0,2)}(\ulq)$. The dotted straight and wavy arrows respectively indicate the amputated propagators $G_{\phi 0}$ and $G_{v 0}$. \label{fig:vertexfun}}
    \end{figure}

In the presence of the nonlinear terms and the multiplicative part of the noises in the Langevin equation, $L_{ij}$ and $\eta_0$ are not the kinetic coefficients that govern the dynamics of the average variables; the functions $\Gamma_{\phi 0}^{(0,2)}$ and $\Gamma_{v 0}^{(0,2)}=\bar\Gamma_{v 0}^{(0,2)}{\cal P}^\perp$ in Eqs.~(\ref{kinetic0}) and (\ref{eta0_vertex}) must be replaced by the two-point vertex functions $\Gamma_{\phi}^{(0,2)}$ and $\bar\Gamma_{v}^{(0,2)}$, respectively, which include the effects of nonlinear coupling of the fluctuations\cite{deDominicis1978,Tauber}. In Appendix \ref{sec:ver_eff}, for the readers unfamiliar with vertex functions, etc., we demonstrate how the effective coefficients are related to the vertex functions in the case of one dynamical variable. The velocity vertex function (tensor) has the form of $\Gamma_{v}^{(0,2)}=\bar\Gamma_{v}^{(0,2)}{\cal P}^\perp$, owing to the isotropy of the system \cite{Robertson1940}. As in the standard field theories\cite{Amit}, the vertex functions are represented by one-particle-irreducible(1PI) graphs with the external propagators being amputated\cite{Tauber}. The two-point vertex functions $\Gamma_{\phi}^{(0,2)}$ and $\Gamma_{v}^{(0,2)}=\bar\Gamma_{v}^{(0,2)}{\cal P}^\perp$ are written as
\begin{align}
    &\Gamma_{\phi}^{(0,2)}(\ulq)=\Gamma_{\phi 0}^{(0,2)}(\ulq)-\Sigma_{\phi}^{(0,2)}(\ulq) \\
    &\Gamma_{v}^{(0,2)}(\ulq)=\Gamma_{v 0}^{(0,2)}(\ulq)-\Sigma_{v}^{(0,2)}(\ulq) \label{vertex_v02}
\end{align}
where $\Sigma_{\phi}^{(0,2)}$ and $\Sigma_{v}^{(0,2)}=\bar\Sigma_{v}^{(0,2)}{\cal P}^\perp$ are called ``self energies" in the terminology of quantum field theory. Figure \ref{fig:vertexfun} shows the general structures of the graphs of the self energies. In the figure, the dotted lines are amputated propagators and the blobs indicate the 1PI graphs. While $\Gamma_{\phi 0}^{(0,2)}$ and $\bar\Gamma_{v 0}^{(0,2)}$ are independent of the frequency $\omega$, the self energies are generally $\omega$-dependent, which indicates the frequency dependence of the kinetic coefficients.
The effective kinetic coefficients ${\cal L}^{\rm eff}_{ij}$ associated with $\phi_i$ at zero frequency is given by ${\cal L}^{\rm eff}_{ij}={\cal L}_{ij}+\Delta{\cal L}_{ij}$ with
\begin{align}
    \Delta{\cal L}_{ij}=\frac{1}{2}\frac{\p [\Sigma_{\phi}^{(0,2)}(\ulq)]_{ij}}{\p (q^2)} \Big|_{({\bm q},\omega)=(0,0)}. \label{Leff1}
\end{align}
The effective viscosity $\eta_{\rm eff}=\eta_0+\Delta\eta$ can similarly be expressed in terms of the vertex function or the self energy. Since we are also interested in the frequency dependence, we keep $\omega$ finite and the excess part $\Delta\eta$ becomes complex,
\begin{align}
    \Delta \eta(\omega)=\Delta \eta'(\omega)-i\Delta \eta''(\omega),
\end{align}
where $\Delta \eta'$ and $-\Delta \eta''$ are respectively the real and imaginary parts.
In terms of the self-energy, the real part is expressed as
\begin{align}
    \Delta\eta'(\omega)=(\rho^2/2T) \frac{\p \bar\Sigma_v^{(0,2)}(\ulq)}{\p (q^2)} \Big|_{q=0}. \label{eta_eff}
\end{align}
Note that the imaginary part $\Delta\eta''$ is determined from $\Delta \eta'$ via the Kramers-Kronig relation. We shall explicitly calculate Eqs.~(\ref{Leff1}) and (\ref{eta_eff}) at the one-loop level in Secs.~\ref{sec:Lij} and \ref{sec:vis}, respectively.

\subsection{Frequency-dependent electrical conductivity \label{sec:conduc0}}
Let us apply an time-dependent, but spatially homogeneous electric field ${\bm E}_0(t)$. 
The dynamic equations are given by the ones in Eqs.~(\ref{deq1}), (\ref{deq2}), and (\ref{deqv}) with $\nabla U$ being replaced by $\nabla U-(e/T){\bm E}_0$. Equation (\ref{deq2}) implies that the charge flux ${\bm J}_{\rm e}$ is given by
\begin{align}
    {\bm J}_{\rm e}/e=& {\bm X}-D[\nabla\phi_2+2\bar n\nabla U+{\bm Y}_1]-\gamma D[\nabla\phi_1+{\bm Y}_2] +(eD/T)[\phi_1+\gamma\phi_2+2\bar n]{\bm E}_0, \label{Je}
\end{align}
where we have defined the composite variables ${\bm X}$, ${\bm Y}_1$, and ${\bm Y}_2$ as
\begin{align}
    &{\bm X}({\bm r},t)\equiv {\bm v}({\bm r},t)\phi_2({\bm r},t) \label{def:X}\\
    &{\bm Y}_i({\bm r},t) \equiv \phi_i({\bm r},t)\nabla U({\bm r},t) \quad (i=1,2).
\end{align}
The action $S^E$ in the presence of ${\bm E}_0$ is also given by the one in Eq.~(\ref{action}) with $\nabla U$ being replaced by $\nabla U-(e/T){\bm E}_0$.

We assume a weak applied field throughout the paper, and hence consider the electrical conductivity tensor $\tensor\sigma(t)$ defined via
\begin{align}
   \av{{\bm J}_{{\rm e}}({\bm r},t)}=\int_{-\infty}^t dt' \tensor\sigma(t-t') {\bm E}_0(t')+O(E_0^2), \label{cond}
\end{align}
where the average in the left hand side is taken with respect to the action $S^E$.
We can rewrite Eq.~(\ref{cond}) in the Fourier domain as
\begin{align}
    \sigma_{\alpha\beta}(\omega)(2\pi)^3\delta_{\bm q}\delta_{\omega -\omega '}= \frac{\delta\av{J_{{\rm e}\alpha}({\bm q},\omega)}}{\delta E_{0\beta}(\omega')}\Big|_{E_0=0}.
\end{align}
We hence have
\begin{align}
    \sigma_{\alpha\beta}(\omega)(2\pi)^3\delta_{\bm q}\delta_{\omega -\omega '}=&\bav{\frac{\delta J_{{\rm e}\alpha}(\ulq)}{\delta E_{0\beta}(\omega')}} \Big|_{E_0=0}-\bav{J_{{\rm e}\alpha}(\ulq)\frac{\delta S^E}{\delta E_{0\beta}(\omega')}}\Big|_{E_0=0}. \label{cond2}
\end{align}
The Fourier transform of Eq.~(\ref{Je}) reads
\begin{align}
    J_{{\rm e}\alpha}(\ulq)/e=&X_\alpha-D[(iq_\alpha/q^2)(q^2+\kappa^2)\phi_2+Y_{1\alpha}] -\gamma D[iq_\alpha\phi_1+Y_{2\alpha}] \nn\\
&+(2eD\bar n/T)(2\pi)^3\delta({\bm q})E_{0\alpha}(\omega)\nn\\
&+\frac{eD}{T}\int_{\omega_1} [\phi_1({\bm q},\omega-\omega_1)+\gamma\phi_2({\bm q},\omega-\omega_1)] E_{0\alpha}(\omega_1), \label{Je2}
\end{align}
where we have used Eq.~(\ref{Poisson3}).
 Using Eq.~(\ref{Je2}) we can readily calculate the first term of Eq.~(\ref{cond2}) as
\begin{align}
    \bav{\frac{\delta J_{{\rm e}\alpha}({\bm q},\omega)}{\delta E_{0\beta}(\omega')}}_0= (2\pi)^3 \delta_{\bm q}\delta_{\omega -\omega '} \delta_{\alpha\beta} 2e^2D\bar n/T.\label{cond3}
\end{align}
If we neglect the second term of Eq.~(\ref{cond2}), we obtain $\sigma_{\alpha\beta}\simeq \sigma^{\rm NE}_{\rm B}\delta_{\alpha\beta}$, where $\sigma^{\rm NE}_{\rm B}$ is the Nernst-Einstein expression for the electrical conductivity,
\begin{align}
    \sigma^{\rm NE}_{\rm B}=2e^2D\bar n/T=e^2(D_++D_-)\bar n/T. \label{sigma_NE}
\end{align}
Here, the subscript ``B'' stands for ``bare'' conductivity because $\sigma^{\rm NE}_{\rm B}$ is written in terms of the bare diffusion coefficient $D$ (See also Secs.~\ref{sec:renormalization} and \ref{sec:electrophoretic}).

In Eq.~(\ref{cond2}), we need to calculate $\delta S^E/\delta E_{0\beta}(\omega)$ from Eqs.~(\ref{action}). For convenience, we introduce the two composite variables,
\begin{align}
    &V_\beta({\bm r},t)\equiv \tilde v_\beta ({\bm r},t)\phi_2({\bm r},t) \label{def:V}\\
    &W_{ij\beta}({\bm r},t)\equiv \phi_i({\bm r},t)\p_\beta \tilde\phi_j({\bm r},t). \label{def:W}
\end{align}
Replacing $\nabla U$ by $\nabla U-(e/T){\bm E}_0$ in Eq.~(\ref{action}), we obtain
\begin{align}
    \frac{\delta (2\pi S^E)}{\delta E_{0\beta}(\omega)}=&-\frac{De}{T}\big[ \gamma \{ W_{11\beta}(0,-\omega)+W_{22\beta}(0,-\omega)\} +W_{12\beta}(0,-\omega)+W_{21\beta}(0,-\omega \big] \nn \\
    &-(e/\rho)V_\beta(0,-\omega). \label{dSdE}
\end{align}
Here $V_\beta(0,-\omega)$ and $W_{ij\beta}(0,-\omega)$ are the Fourier-transformed composite variables evaluated at $({\bm q}=0, -\omega)$. Substitution of Eqs.~(\ref{Je2}) and (\ref{dSdE}) into (\ref{cond2}) yields the full expression for the electrical conductivity, which shall be calculated at the one-loop level in Sec.~\ref{sec:conduc}.

\section{Calculation and Results\label{sec:result}}
In this section, we investigate the effective kinetic coefficients ${\cal L}^{\rm eff}_{ij}$, the frequency-dependent viscosity $\eta_{\rm eff}(\omega)$, and the frequency-dependent electrical conductivity $\sigma_{\alpha\beta}(\omega)$ from Eqs.~(\ref{Leff1}), (\ref{eta_eff}) and (\ref{cond2}) within the one-loop approximation. For the sake of lighter notation, the components of the propagators are hereafter indicated by superscripts:
\begin{align}
    &G_{\phi 0}^{ij}=[G_{\phi 0}]_{ij},\quad G_{v0}^{\alpha\beta}=[G_{v0}]_{\alpha\beta} ,\quad C_{\phi 0}^{ij}=[C_{\phi 0}]_{ij},\quad C_{v0}^{\alpha\beta}=[C_{v0}]_{\alpha\beta} .\nn 
\end{align}
Similarly, we use the following notations: 
\begin{align}
    &\Gamma_{\phi,ij}^{(0,2)}=[\Gamma_{\phi}^{(0,2)}]_{ij},\quad \Gamma_{v,\alpha\beta}^{(0,2)}=[\Gamma_{v}^{(0,2)}]_{\alpha\beta} ,\quad \Sigma_{\phi,ij}^{(0,2)}=[\Sigma_{\phi}^{(0,2)}]_{ij},\quad \Sigma_{v,\alpha\beta}^{(0,2)}=[\Sigma_{v}^{(0,2)}]_{\alpha\beta}. \nn
\end{align}
We also define the Debye wavenumber,
\begin{align}
    \kappa=\sqrt{8\pi \ell_{\rm B} \bar n}.
\end{align}

The explicit expressions for propagators are readily obtained from Eqs.~(\ref{Gamma_p0_1})--(\ref{Gamma_v0_1}) and (\ref{G_p0})--(\ref{C_v0}):
\begin{align}
&G_{\phi 0}(\ulq )=\frac{1}{(i\omega +\lambda_1)(i\omega+\lambda_2)}\begin{pmatrix}
i\omega+D(q^2+\kappa^2) & -\gamma D(q^2+\kappa^2) \\
-\gamma Dq^2 & i\omega +Dq^2 
\end{pmatrix} \label{Gp0}\\
&C_{\phi 0}(\ulq )=\frac{4\bar n Dq^2}{(\omega^2 +\lambda_1^2)(\omega^2+\lambda_2^2)} \begin{pmatrix}
\omega^2 +D^2(1-\gamma^2)(q^2+\kappa^2)^2 & \gamma [\omega^2-\lambda_1\lambda_2] \\
 \gamma [\omega^2-\lambda_1\lambda_2]& \omega^2 +D^2(1-\gamma^2)q^4
\end{pmatrix} \label{Cp0} \\
    & G_{v0}^{\alpha\beta} (\ulq )=\frac{{\cal P}^\perp_{\alpha\beta} ({\bm q})}{i\omega+\nu_0 q^2} \\
    &C_{v0}^{\alpha\beta} (\ulq )=2T(\eta_0/\rho^2)q^2 \frac{{\cal P}^\perp_{\alpha\beta} ({\bm q})}{\omega^2+\nu_0^2 q^4}, \label{Prop_v}
\end{align}
where $\lambda_1(q)$ and $\lambda_2(q)$ are defined via ${\rm det}\, \Gamma_{\phi 0}^{(1,1)}(\ulq)=(i\omega+\lambda_1)(i\omega+\lambda_2)$ yielding
\begin{align}
    \lambda_1+\lambda_2=D(2q^2+\kappa^2),\quad \lambda_1\lambda_2=D^2(1-\gamma^2)q^2(q^2+\kappa^2). \label{eigen}
\end{align}
Note that for a symmetric salt, i.e., $\gamma=0$, the matrices $G_{\phi 0}$ and $C_{\phi 0}$ become diagonal, and we have $\lambda_1=Dq^2$ and $\lambda_2=D(q^2+\kappa^2)$.

\subsection{Effective kinetic coefficients for density variables and renormalization of diffusion coefficients\label{sec:Lij}}
\begin{figure}[htp]
    \includegraphics[width=0.8\columnwidth]{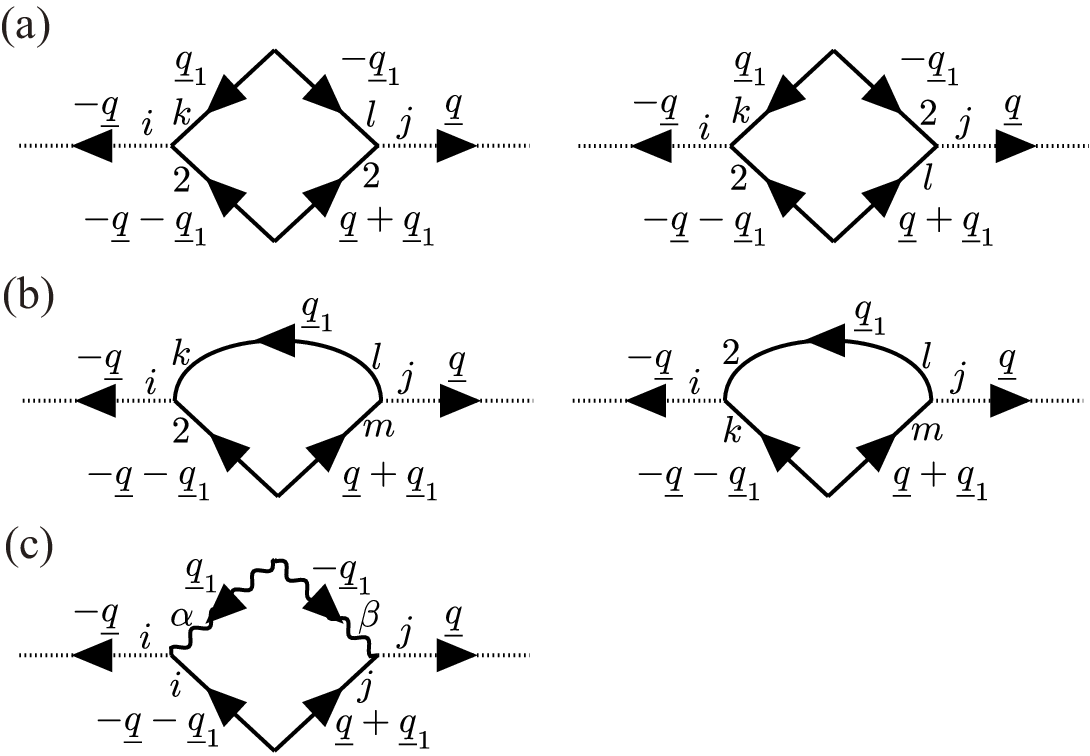}
    \caption{One-loop graphs of the self-energy $\Sigma_{\phi,ij}^{(0,2)}(\ulq)$ associated with the density variables. The dotted lines represent amputated propagators $G_{\phi 0}$. Note that the indices $k$ and $l$ in (a),  $k$, $l$, and $m$ in (b), and $\alpha$ and $\beta$ in (c) are summed over. Note also that there are two other graphs in (b) in which $i$ and $j$ are exchanged.\label{fig:L}}
\end{figure}

We first discuss the effective kinetic coefficients for the density variables  ${\cal L}_{ij}^{\rm eff}$ (or $L_{ij}^{\rm eff}$), which is the counterparts of ${\cal L}_{ij}$ (or $L_{ij}$) accounting for the fluctuation effects. We shall see that they also lead to the renormalization of the diffusion coefficients introduced recently\cite{Peraud2017,Donev2019}.
Figure \ref{fig:L}(b) presents the one-loop graphs for the vertex function $\Gamma_{\phi,ij}^{(0,2)}({\bm q},\omega)$, or, the self energy $\Sigma_{\phi,ij}^{(0,2)}({\bm q},\omega)$. Each graph in (a) consists of two $S_{\phi\phi}$-vertices (See Fig.~\ref{fig:vertices}a). Similarly, each of (b) is made of one $S_{\phi\phi}$-vertex and one $S_{\rm m}$-vertex, and the graph (c) is of two $S_{\phi v}$-vertices. To be precise, there are two other graphs in (b) where $i$ and $j$ are exchanged. We are interested in the slowly varying, long-wavelength modes, so we set $\omega=0$ and expand $\Sigma_{\phi,ij}^{(0,2)}({\bm q},0)$ in powers of $q^2$; the coefficient of $q^2$ gives the correction to the kinetic coefficient $\Delta{\cal L}_{ij}$ [See Eq.(\ref{Leff1})].

For convenience we divide $\Delta{\cal L}_{ij}$ into three parts $\Delta{\cal L}_{ij}^{\rm a}$, $\Delta{\cal L}_{ij}^{\rm b}$, and $\Delta{\cal L}_{ij}^{\rm c}$, which are respectively the contributions from Fig.~\ref{fig:L}a, b, and c:
\begin{align}
    \Delta{\cal L}_{ij}=\Delta{\cal L}_{ij}^{\rm a}+\Delta{\cal L}_{ij}^{\rm b}+\Delta{\cal L}_{ij}^{\rm c}.
\end{align} 

\subsubsection{Contribution from the graphs in Figs.~\ref{fig:L}a \label{sec:diff1}}
First we consider the two graphs in Figs.~\ref{fig:L}a.  As shown in Appendix \ref{sec:Lija}, the terms proportional to $q^2$ yield
\begin{align}
    \Delta{\cal L}_{11}^{\rm a}=(4\pi \ell_{\rm B}\gamma D)^2 \int _{\ulq_1} \Big[&C_{\phi 0}^{22}(\ulq_1)C_{\phi 0}^{11}(\ulq_1)-C_{\phi 0}^{12}(\ulq_1)^2\Big] \frac{(\hat{\bm q}_1\cdot \hat{\bm q})^2}{2q_1^2}, \label{L11a0}
\end{align}
where the bold symbols with a caret $\hat{}$ denote the normalized wave vectors, e.g., $\hat {\bm q}={\bm q}/q$. Substituting Eq.~(\ref{Cp0}) into the above expression and performing the $\omega_1$-integration, we obtain
\begin{align}
    \Delta{\cal L}_{11}^{\rm a}=&(4\pi \ell_{\rm B}\gamma D)^2 \int _{q_1} \frac{4\bar n^2 D^2 q_1^2 (1-\gamma^2)}{\lambda_1\lambda_2(\lambda_1+\lambda_2)}(\hat{\bm q}_1\cdot \hat{\bm q})^2 
    =\gamma^2 D\kappa^3 \frac{2-\sqrt{2}}{24\pi} \label{L11a}
\end{align}
where $\int_{q_1}=(2\pi)^{-3}\int d{\bm q}_1$, and use has been made of Eq.~(\ref{eigen}). The other two terms $\Delta{\cal L}_{22}^{\rm a}$ and $\Delta{\cal L}_{12}^{\rm a}$ differ from $\Delta{\cal L}_{11}^{\rm a}$ only by the multiplicative factors $\gamma^{-2}$ and $\gamma^{-1}$, respectively (see Appendix \ref{sec:Lija}):
\begin{align}
    &\Delta{\cal L}_{22}^{\rm a}= D\kappa^3 \frac{2-\sqrt{2}}{24\pi}\label{L22a}\\
    &\Delta{\cal L}_{12}^{\rm a}=\Delta{\cal L}_{21}^{\rm a}= \gamma D\kappa^3 \frac{2-\sqrt{2}}{24\pi}.\label{L12a}
\end{align}

\subsubsection{Contribution from the graphs in Figs.~\ref{fig:L}b}
As shown in Appendix \ref{sec:Lijb}, the graphs in Figs.~\ref{fig:L}b yield for $i=j=1$
\begin{align}
    \Delta{\cal L}_{11}^{\rm b}=&-2\gamma^2D\kappa^4\int_{q_1} \frac{(\hat{\bm q}\cdot \hat{\bm q}_1)^2}{(2q_1^2+\kappa^2)(q_1^2+\kappa^2)} 
    = -\gamma^2 D\kappa^3 \frac{2-\sqrt{2}}{12\pi}, \label{L11b}
\end{align}
and similarly for $i=j=2$
\begin{align}
    \Delta{\cal L}_{22}^{\rm b}=-D\kappa^3 \frac{2-\sqrt{2}}{12\pi}.\label{L22b}
\end{align}
For $(i,j)=(1,2)$ and (2,1), one can easily find $\Delta{\cal L}_{12}^{\rm b}=\Delta{\cal L}_{21}^{\rm b}=(\Delta{\cal L}_{11}^{\rm b}/\gamma+\gamma\Delta{\cal L}_{22}^{\rm b})/2$ using Eqs.~(\ref{vertex_u1_2})--(\ref{v1i_2}). We hence have
\begin{align}
    \Delta{\cal L}_{12}^{\rm b}=\Delta{\cal L}_{21}^{\rm b}=-\gamma D\kappa^3 \frac{2-\sqrt{2}}{12\pi}.\label{L12b}
\end{align}

\subsubsection{Contribution from the graph in Fig.~\ref{fig:L}c}
The analytical expression of the graph in Fig.~\ref{fig:L}c is
\begin{align}
    \mbox{Fig.~\ref{fig:L}c}=\sum_{\alpha,\beta} \int_{\ulq_1}& C_{v0}^{\alpha\beta}(\ulq_1)C_{\phi 0}^{ij}(\ulq_1+\ulq)  s_\alpha({\bm q}_1,-{\bm q}_1-{\bm q}) s_\beta (-{\bm q}_1,{\bm q}_1+{\bm q}). \label{diff_v}
\end{align}
We first take the sum over the indices $\alpha$ and $\beta$. From Eqs.~(\ref{vertex_s}) and (\ref{Prop_v}) we obtain
\begin{align}
    \sum_{\alpha,\beta} s_\alpha({\bm q}_1,-{\bm q}_1-{\bm q}) s_\beta (-{\bm q}_1,{\bm q}_1+{\bm q}) C_{v0}^{\alpha\beta}(\ulq_1) 
    = \bar C_{v0}(\ulq_1) q^2\{ 1-(\hat{\bm q}_1\cdot\hat{\bm q})^2\}, \label{Lc_factor}
\end{align}
where $\bar C_{v0}(\ulq_1)=2(T\eta_0/\rho^2)q_1^2/(\omega_1^2+\nu_0^2q_1^4)$. The factor Eq.~(\ref{Lc_factor}) is of order $q^2$, so we have
\begin{align}
    \Delta {\cal L}_{ij}^{\rm c}=\int_{\ulq_1} \bar C_{v0}(\ulq_1)C_{\phi 0}^{ij}(\ulq_1) \{ 1-(\hat{\bm q}_1\cdot\hat{\bm q})^2\}. \label{L11c0}
\end{align}

Meanwhile, the typical values of the kinetic viscosity $\nu_0$ and the (mean) ion diffusion coefficient $D$ are $10^{-6}\, {\rm m}^2/{\rm s}$ and $10^{-9}\, {\rm m}^2/{\rm s}$, respectively, for aqueous electrolyte solutions.
Hence we hereafter assume that the Schmidt number is large:
\begin{align}
    \hat\nu_0\equiv \nu_0/D \gg 1. \label{schmidt}
\end{align}
To be precise, the diffusion coefficient $D$ in the above should be replaced by the renormalized coefficient $D_{\rm R}$ introduced later in Eq.~(\ref{D_R}).
For $i=j=1$, substituting Eq.~(\ref{Cp0}) into Eq.~(\ref{L11c0}) and performing $\omega_1$-integration, we obtain
\begin{align}
    \Delta {\cal L}_{11}^{\rm c}=\frac{2\bar n T}{3\pi ^2\eta_0}\int dq_1,
\end{align} 
where use has been made of Eq.~(\ref{schmidt}).
To regularize the ultraviolet divergence of the above integral, we introduce an upper cut-off wavenumber $\Lambda$ such that $2\pi\Lambda^{-1}$ is comparable to the molecular size. We then obtain
\begin{align}
    \Delta {\cal L}_{11}^{\rm c}= \frac{2\bar n T\Lambda}{3\pi^2\eta_0}. \label{L11c}
\end{align}
Similarly we obtain
\begin{align}
    &\Delta {\cal L}_{22}^{\rm c}
    =\frac{2\bar n T}{3\pi^2\eta_0}\Big[ \Lambda-\frac{\pi\kappa}{2}\Big]. \label{L22c} \\
    &\Delta {\cal L}_{12}^{\rm c}=\Delta {\cal L}_{21}^{\rm c}
    =  -\frac{2\gamma\bar n T}{3\pi^2\eta_0} \Big[ \Lambda \hat\nu_0^{-1}-\frac{\pi\kappa}{2}\hat\nu_0^{-3/2}\Big]. \label{L12c}
\end{align}

\subsubsection{Total effective coefficients and Renormalization of diffusion coefficients\label{sec:renormalization}}
Collecting Eqs.~(\ref{Lij_0loop}), (\ref{L11a})--(\ref{L12a}), (\ref{L11b})--(\ref{L12b}), and (\ref{L11c})--(\ref{L12c}), we obtain the effective kinetic coefficients at the one-loop approximation,
\begin{align}
    &{\cal L}_{11}^{\rm eff}=2\bar n D -\gamma^2 D\kappa^3 \frac{2-\sqrt{2}}{24\pi}+ \frac{2\bar n T\Lambda}{3\pi^2\eta_0} \label{Leff11}\\
    &{\cal L}_{22}^{\rm eff}=2\bar n D - D\kappa^3 \frac{2-\sqrt{2}}{24\pi}+\frac{2\bar n T}{3\pi^2\eta_0}\Big[ \Lambda-\frac{\pi\kappa}{2}\Big] \\
    &{\cal L}_{12}^{\rm eff}=2\bar n \gamma D -\gamma D\kappa^3 \frac{2-\sqrt{2}}{24\pi}-\frac{2\gamma\bar n T}{3\pi^2\eta_0}\Lambda \hat\nu_0^{-1}. \label{Leff12}
\end{align}
Here, in the last line, we have neglected the term proportional to $(D/\nu_0)^{3/2}$ appearing in Eq.~(\ref{L12c}), because this term is much smaller than the second term of Eq.~(\ref{Leff12}); we can estimate the ratio between these terms as
\begin{align}
    \frac{T}{\eta_0\ell_{\rm B} D}\hat\nu_0^{-3/2} \sim 6\hat\nu_0^{-3/2}  \ll 1
\end{align}
for electrolyte solutions in ambient conditions where $D\sim 10^{-9}\, {\rm m}^2/{\rm s}$, $\eta_0\sim 10^{-3}\, {\rm Pa}\cdot{\rm s}$, and $\ell_{\rm B}\approx 7$\AA  [As in Eq.~(\ref{schmidt}), $D$ in this argument should be replaced by the renormalized coefficient $D_{\rm R}$ which is to be introduced in Eq.~(\ref{D_R})].

The $\Lambda$-sensitive terms in Eqs.~(\ref{Leff11})--(\ref{Leff12}) can be absorbed into the renormalized mean diffusion coefficients $D_{\rm R}$ and the asymmetry coefficient $\gamma_{\rm R}$. This observation has been pointed out previously\cite{Peraud2017,Donev2019}. Note that the renormalization of the diffusion coefficient due to the fluid velocity is not unique to electrolyte solutions, but is necessary for non-ionic solutions \cite{Bedeaux1974,Mazur1974}.
At the one-loop level, they are given by 
\begin{align}
    &D_{\rm R}= D+\frac{T\Lambda}{3\pi^2\eta_0} \label{D_R}\\
    &\gamma_{\rm R}=\gamma \Big[1-\frac{T\Lambda}{3\pi^2 \eta_0 } \Big(\frac{1}{\nu_0}+\frac{1}{D}\Big)\Big].
\end{align}
One can interpret the second term of Eq.~(\ref{D_R}) as the contribution analogous to the Stokes-Einstein diffusivity which stems from the drag force on the ions due to the fluid flow; one obtains $T\Lambda/(3\pi^2\eta_0)=T/(6\pi\eta_0 a)$ with $a$ being the length comparable with the ionic radii if setting $\Lambda=\pi/2a$. However, the equality $\Lambda=\pi/2a$ should not be taken too seriously for the following two reasons. (i) Strictly speaking, the Stokes-Einstein diffusivity cannot be applied to such small objects as ions (although the analogous drag force on the ions due to the fluid flow should exist). (ii)Furthermore, the cut-off wavenumber $\Lambda$, up to which the continuum description (Diffusion equation and Stokes equation) are assumed to be valid, should be considerably smaller than $a^{-1}$.

In terms of these renormalized coefficients, we can rewrite Eqs.~(\ref{Leff11})--(\ref{Leff12}) as
\begin{align}
    &{\cal L}_{11}^{\rm eff}=2D_{\rm R}\bar n-\gamma_{\rm R}^2 D_{\rm R}\kappa^3 \frac{2-\sqrt{2}}{24\pi} \\
    &{\cal L}_{22}^{\rm eff}=2D_{\rm R}\bar n-D_{\rm R}\kappa^3 \frac{2-\sqrt{2}}{24\pi}-\frac{T\bar n \kappa}{3\pi\eta_0} \\
    &{\cal L}_{12}^{\rm eff}=2\gamma_{\rm R} \Big[ D_{\rm R}\bar n-D_{\rm R}\kappa^3 \frac{2-\sqrt{2}}{48\pi}\Big],
\end{align}
where in the terms of one-loop corrections we have replaced the bare coefficients $D$ and $\gamma$ by the renormalized ones in the logic of perturbation theory (the error due to the replacement is of the two-loop order).
We can convert these expressions associated to $\phi_i$ into those associated to $n_+$ and $n_-$ using the trivial relations $L_{++}^{\rm eff}=({\cal L}_{11}^{\rm eff}+{\cal L}_{22}^{\rm eff}+2{\cal L}_{12}^{\rm eff})/4$, $L_{--}^{\rm eff}=({\cal L}_{11}^{\rm eff}+{\cal L}_{22}^{\rm eff}-2{\cal L}_{12}^{\rm eff})/4$ and $L_{+-}^{\rm eff}=({\cal L}_{11}^{\rm eff}-{\cal L}_{22}^{\rm eff})/4$,
\begin{align}
    &L_{++}^{\rm eff}=D_{\rm +R}\bar n -\frac{D_{\rm +R}^2\kappa^3(2-\sqrt{2})}{48\pi (D_{\rm +R}+D_{\rm -R})}-\frac{T\bar n \kappa}{12\pi \eta_0} \label{Lpp_eff}\\
    &L_{--}^{\rm eff}=D_{\rm -R}\bar n -\frac{D_{\rm -R}^2\kappa^3(2-\sqrt{2})}{48\pi (D_{\rm +R}+D_{\rm -R})}-\frac{T\bar n \kappa}{12\pi \eta_0} \label{Lmm_eff}\\
    &L_{+-}^{\rm eff}=\frac{D_{\rm +R}D_{\rm -R}\kappa^3(2-\sqrt{2})}{48\pi (D_{\rm +R}+D_{\rm -R})}+\frac{T\bar n \kappa}{12\pi \eta_0}. \label{Lpm_eff}
\end{align}
The coefficients $D_{\rm +R}$ and $D_{\rm -R}$ are the renormalized diffusion coefficients for cations and anions, respectively, which are given by 
\begin{align}
    D_{\rm \pm R}=(1\pm\gamma_{\rm R})D_{\rm R} =D_\pm +\frac{T\Lambda}{3\pi^2\eta_0}\Big[1\mp \gamma \hat\nu_0^{-1}\Big].
\end{align}
We notice that the cross-kinetic coefficient $L_{+-}^{\rm eff}$, which is absent in the starting Langevin equation, appears as a result of the nonlinear interaction of fluctuations. Previously, Eqs.~(\ref{Lpp_eff})--(\ref{Lpm_eff}) for symmetric salts $D_{\rm +R}=D_{\rm -R}$ have been anticipated from the electrical conductivity at zero-frequency\cite{Peraud2017}, but in the present study we have derived them  calculating ``blindly" the associated vertex functions. 

\subsection{Frequency-dependent shear viscosity\label{sec:vis}}
\begin{figure}[htp]
    \includegraphics[width=0.4\columnwidth]{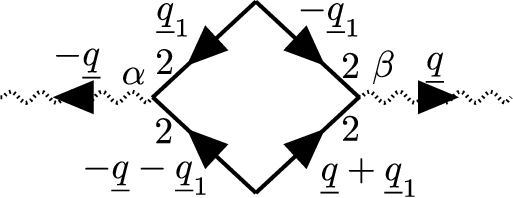}
    \caption{One-loop graphs of the self energies $\Sigma_{v,\alpha\beta}^{(0,2)}(\ulq)$. The dotted lines represent amputated propagators $G_{v0}$.\label{fig:graph_vis}}
\end{figure}

We next discuss the effective shear viscosity. Figure \ref{fig:graph_vis} shows the one-loop graph of the fluctuation correction to the vertex function $\Gamma_v^{(0,2)}$, i.e., the self-energy $\Sigma_v^{(0,2)}$. Its analytical expression is
\begin{align}
    \Sigma_{v,\alpha\beta}^{(0,2)}(\ulq)=-\frac{1}{2}\int_{\ulq_1} &C_{\phi 0}^{22}(\ulq_1)C_{\phi 0}^{22}(\ulq_1+\ulq)  w_\alpha(-{\bm q}_1,{\bm q}_1+{\bm q})w_\beta(-{\bm q}_1,{\bm q}_1+{\bm q}), \label{self_v}
\end{align}
where the vertex factor $w_\alpha$ has been defined in Eq.~(\ref{vertex_w}).

Meanwhile, using Eq.~(\ref{vertex_w}), we obtain to the order of $q^2$,
\begin{align}
    -\sum_\alpha w_\alpha(-{\bm q}_1,{\bm q}_1+{\bm q})^2 = 64\pi^2\ell_{\rm B}^2(T/\rho)^2 q^2q_1^{-4}(\hat{\bm q}\cdot\hat{\bm q}_1)^2[1-(\hat{\bm q}\cdot\hat{\bm q}_1)^2]. \label{sum_ww}
\end{align}
As mentioned around Eq.~(\ref{vertex_v02}), the self-energy has the from $\Sigma_{v,\alpha\beta}^{(0,2)}(\ulq)=\bar\Sigma_{v}^{(0,2)}(\ulq){\cal P}_{\alpha\beta}^\perp$, so we have $\bar\Sigma_{v}^{(0,2)}(\ulq)=(1/2)\sum_\alpha \Sigma_{v,\alpha\alpha}^{(0,2)}(\ulq)$.
Hence from Eqs.~(\ref{eta_eff}), (\ref{self_v}), and (\ref{sum_ww}) we obtain
\begin{align}
    \Delta\eta'=8\pi^2\ell_{\rm B}^2T \int_{\ulq_1} &C_{\phi 0}^{22}({\bm q}_1,\omega_1)C_{\phi 0}^{22}({\bm q}_1,\omega_1+\omega)  q_1^{-4}(\hat{\bm q}\cdot\hat{\bm q}_1)^2[1-(\hat{\bm q}\cdot\hat{\bm q}_1)^2].
    \label{self_v2}
\end{align}
Here we introduce the dimensionless frequency,
\begin{align}
    \Omega\equiv \omega/D\kappa^2, \label{dless0}
\end{align}
where $D\kappa^2$ is the characteristic frequency of the relaxation of the ionic atmosphere. Accordingly, in the integral of Eq.~(\ref{self_v2}), we define
\begin{align}
s\equiv q_1/\kappa,\quad \hat \lambda_i(s)\equiv\lambda_i(q_1)/D\kappa^2, \label{dless} 
\end{align}
which are also dimensionless. Performing the frequency-integration and the angular-integration (see Appendix \ref{deriv_H}), we obtain
\begin{align}
    \Delta\eta'=\frac{2T\kappa}{15\pi^2 D_{\rm R}}H(\Omega^2;\gamma_{\rm R}^2), \label{eta_excess}
\end{align}
where, again, $D$ and $\gamma$ have been replaced by $D_{\rm R}$ and $\gamma_{\rm R}$ in the logic of perturbation theory [To be precise, we have also implicitly made the replacement $D\to D_{\rm R}$ in Eqs.~(\ref{dless0}) and (\ref{dless})]. The function $H$ is given by
\begin{align}
    H(\Omega^2;\gamma^2)= 
    \int_0^\infty ds\, &\frac{s^2}{(s^2+1)(4\hlamb_1^2+\Omega^2)(4\hlamb_2^2+\Omega^2)\{(2s^2+1)^2+\Omega^2\}} \nn\\
    &\times\big[\Omega^4+\Omega^2\{ (2s^2+1)(5s^2+1)-(1-\gamma^2)(2s^4+3s^2)\}\nn\\
    &\hspace{6mm}-4(1-\gamma^2)s^4(2s^2+1)\{\gamma^2(s^2+1)-(2s^2+1)\} 
    \big]. \label{H_scaling}
\end{align}
We notice that, at the one-loop level, the frequency dependence is determined by the ratio $\Omega=\omega/D_{\rm R}\kappa^2$, which means that the dispersion is dominated by the slow relaxation of ionic atmosphere distorted by the fluid flow. We could have anticipated this observation from the one-loop vertex graph in Fig.~\ref{fig:graph_vis} which involves no velocity propagators (wavy solid lines).

At zero-frequency, $\Omega=0$, Eq.~(\ref{H_scaling}) reduces to
\begin{align}
    H(0;\gamma_{\rm R}^2)=&\int_0^\infty ds\, \frac{s^2[(2s^2+1)-\gamma_{\rm R}^2(s^2+1)]}{4(s^2+1)^3(2s^2+1)(1-\gamma_{\rm R}^2)} 
    =\frac{\pi}{64}\Big[ 1+\frac{\gamma_{\rm R}^2}{1-\gamma_{\rm R}^2} (8\sqrt{2}-11)\Big],
\end{align}
which reproduces the classical result,
\begin{align}
    \eta_{\rm eff}(0)-\eta_0=\frac{T\kappa}{480\pi D_{\rm R}}\Big[ 1+\frac{\gamma_{\rm R}^2}{1-\gamma_{\rm R}^2} (8\sqrt{2}-11)\Big]. \label{vis_eff_0}
\end{align}
This was originally derived by Falkenhagen and Vernon\cite{falkenhagenLXIIViscosityStrong1932}, and later generalized to multicomponent salts by Onsager {\it et al.} \cite{onsagerIrreversibleProcessesElectrolytes1932,OnsagerKim1957}. More recently, Wada have derived the same expression for $\gamma_{\rm R}=0$ using the linearized FHE \cite{Wada2005}.

\begin{figure}[htp]
    \includegraphics[width=0.8\columnwidth]{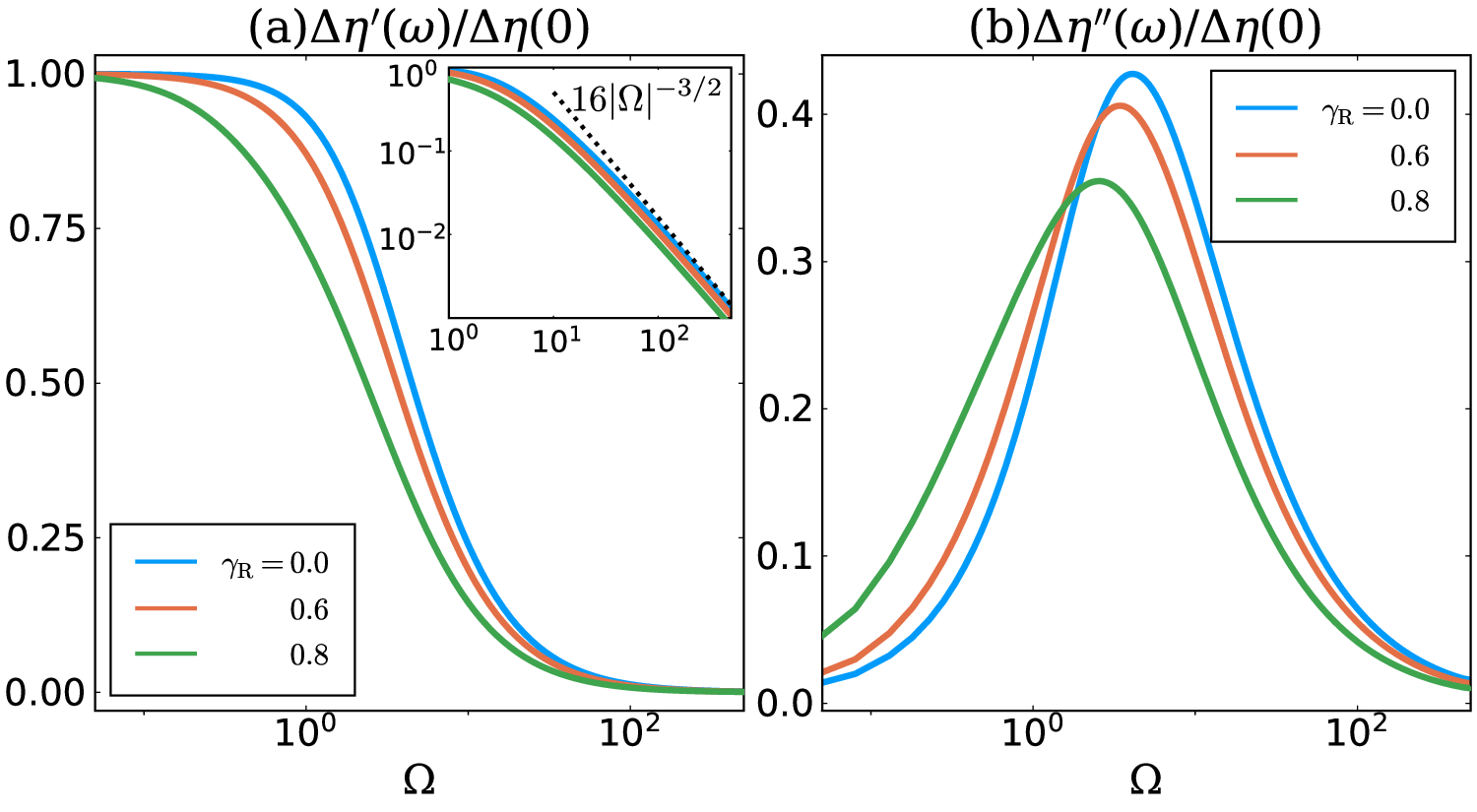}
    \caption{The complex excess viscosity $\Delta \eta(\omega)$ vs the normalized frequency $\Omega=\omega/D_{\rm R}\kappa^2$: (a)real part $\Delta\eta'$ Inset: the same plots in log-log scale. and (b)imaginary part $\Delta\eta''$. \label{fig:eta}}
\end{figure}

For non-zero frequencies, we perform numerical integration of Eq.~(\ref{H_scaling});  Figure \ref{fig:eta}a shows the real part normalized by the zero-frequency value, $\Delta\eta'(\omega)/\Delta\eta(0)=H(\Omega;\gamma_{\rm R})/H(0;\gamma_{\rm R})$ as a function of $\Omega$ for $\gamma_{\rm R}=0$, 0.6, and 0.8. It is almost constant for $1\ll \Omega$, and starts to decay around $\Omega \sim 1$ ($\omega\sim D_{\rm R}\kappa^2$). We also notice that  $\Delta\eta'$ decays faster for larger values of $|\gamma_{\rm R}|$. As $\Omega$ is further increased, $\Delta\eta'$ decays as $\Omega^{-3/2}$, and eventually vanishes for $\Omega \gtrsim  100$. That is, the oscillation of the shear flow is too fast for the ionic atmosphere to relax. In Fig.~\ref{fig:eta}b, the normalized imaginary part $\Delta\eta''(\omega)/\Delta\eta(0)$ is plotted as a function of $\Omega$. Here, $\Delta \eta''$ has been calculated from $\Delta\eta'$ using the Kramers-Kronig relation. In accordance with the dispersion behavior of $\Delta\eta'$, the imaginary part $\Delta\eta''(\omega)$ starts to grow at around $\Omega\sim 1$ and vanishes for $\Omega \gg 100$. 

\begin{figure}[htp]
    \includegraphics[width=0.8\columnwidth]{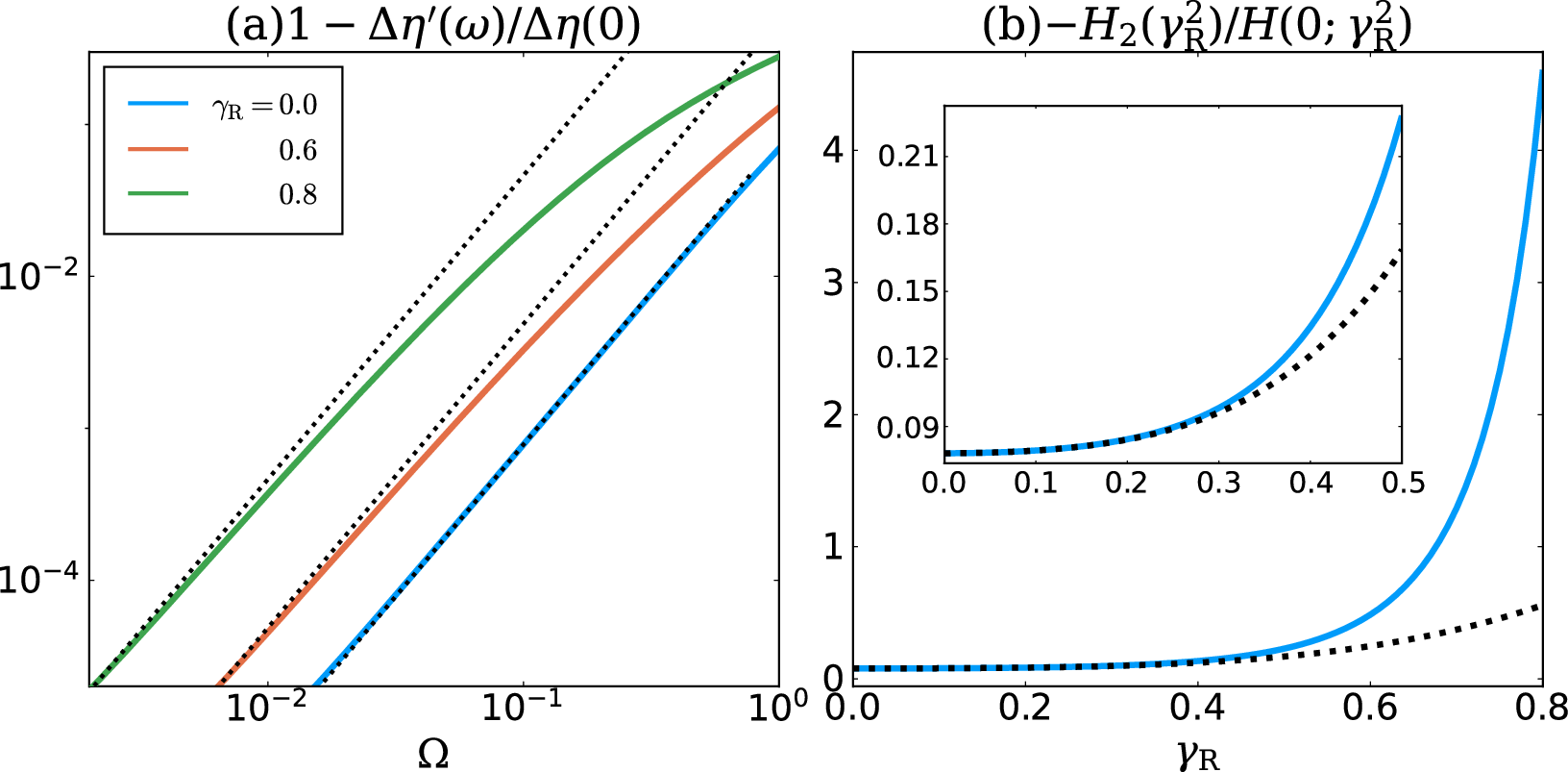}
    \caption{(a) Logarithmic plots of the real part $\Delta \eta'(\omega)$ for small $\Omega=\omega/D_{\rm R}\kappa^2$ at $\gamma_{\rm R}=0.0$, $0.6$, and $0.8$. The dotted lines show the limiting function for $\Omega\to 0$, i.e., $-\Omega^2 H_2(\gamma_{\rm R}^2)/H(0;\gamma_{\rm R}^2)$ . (b)The coefficient $-H_2(\gamma_{\rm R}^2)/H(0;\gamma_{\rm R}^2)$ as a function of $\gamma_{\rm R}$. The dotted line shows the Taylor expansion with respect to $\gamma_{\rm R}^2$ up to the order of $\gamma_{\rm R}^4$. Inset: the same plot in a narrower $\gamma_{\rm R}$-range. \label{fig:eta0}}
\end{figure}

To derive the asymptotic behavior $\Delta\eta' \sim |\Omega|^{-3/2}$ for $|\Omega|\to\infty$, we change the variable $s\to y=s/|\Omega|^{1/2}$ in Eq.~(\ref{H_scaling}). Using Eq.~(\ref{eigen}), we find for $|\Omega|\to \infty$,
\begin{align}
    H(\Omega^2;\gamma_{\rm R}^2) \to |\Omega|^{-3/2} K(\gamma_{\rm R}^2)
\end{align}
with
\begin{align}
    K(\gamma^2) 
    &=\int_0^\infty dy\,\frac{1+2(4+\gamma^2)y^4+8(1-\gamma^2)(2-\gamma^2)y^8}{\{1+8(1+\gamma^2)y^4+16(1-\gamma^2)^2y^8\}(4y^4+1)} \nn\\
    &= \frac{\pi}{4}-\frac{\pi}{64}\gamma^2+O(\gamma^4).
\end{align}
The inset of Fig.~\ref{fig:eta}a presents the log-log scale plots of $\Delta\eta'(\omega)/\Delta\eta(0)$, and the asymptotic curve $|\Omega|^{-3/2}K(0)/H(0;0)=16|\Omega|^{-3/2}$ for $\gamma_{\rm R}=0$. The asymptotic curve is in good agreement with the numerical result for $\Omega \gtrsim 10$.

Conversely, for small $\Omega$, we expand $H(\Omega;\gamma_{\rm R})$ in powers of $\Omega^2$:
\begin{align}
    H(\Omega^2;\gamma_{\rm R}^2)=H(0;\gamma_{\rm R}^2)+H_2(\gamma_{\rm R}^2)\Omega^2+O(\Omega^4).
\end{align}
From Eq.~(\ref{H_scaling}) we obtain the coefficient $H_2$,
\begin{align}
    H_2(\gamma^2)=&\frac{-\pi}{4096(1-\gamma^2)^3} [5-h_1\gamma^2+h_2 \gamma^4-h_3\gamma^6],
\end{align}
where $h_1=2088-1472\sqrt{2}\approx 6.28$, $h_2=4944-3456\sqrt{2}\approx 56.5$, and $h_3=2816-1984\sqrt{2}\approx 10.2$.
We hence have for small $|\Omega|$,
\begin{align}
    1-\frac{\Delta\eta'(\omega)}{\Delta \eta(0)}\simeq &-\frac{H_2(\gamma_{\rm R}^2)}{H(0;\gamma_{\rm R}^2)}\Omega^2 \label{eta_small}\\
    \simeq &\frac{1}{64}\Big[5+a_1\gamma_{\rm R}^2+a_2\gamma_{\rm R}^4\Big]\Omega^2, \label{eta_small2}
\end{align}
where in the second line we have expanded $H_2(\gamma_{\rm R}^2)/H(0;\gamma_{\rm R}^2)$ in powers of $\gamma_{\rm R}^2$ to the second order; the coefficients are given by the positive numbers, $a_1=1432\sqrt{2}-2018\approx 7.15$ and $a_2=32816\sqrt{2}-46345\approx 63.83$.
Figure \ref{fig:eta0}a shows the log-log scale plots of the left hand side of Eq.~(\ref{eta_small}) with solid lines and the right hand side of Eq.~(\ref{eta_small}) with dotted lines for $\gamma_{\rm R}=0$, 0.6, and 0.8, where the left hand side has been calculated numerically from Eq.~(\ref{H_scaling}). It shows that the right hand side of Eq.~(\ref{eta_small}) well describes the initial decay of $\Delta\eta'(\omega)/\Delta\eta(0)$ for $\Omega\lesssim 0.1$. In Fig.~\ref{fig:eta0}b, the decay rate $-H_2(\gamma_{\rm R})/H(0;\gamma_{\rm R}^2)$ is plotted as a function of $\gamma_{\rm R}$. It increases gradually for small $\gamma_{\rm R}$ and grows steeply for $\gamma_{\rm R}\gtrsim 0.6$. This means that for largely asymmetric salts the viscosity $\Delta\eta'$ decays much faster than for moderately asymmetric ones if plotted as a function of the scaled frequency $\Omega$. Hydrochloric acid, HCl, is an example of such a largely asymmetric salt, for which the diffusion coefficients are\cite{RobinsonStokes} $D_{\rm +R}=9.3\times 10^{-9} {\rm m}^2/{\rm s}$ and $D_{\rm -R}=2.0\times 10^{-9} {\rm m}^2/{\rm s}$ yielding $\gamma_{\rm R}=0.65$.
Figure \ref{fig:eta0}b also presents the approximate slope in Eq.~(\ref{eta_small2}) with a dotted line; it is accurate for $\gamma_{\rm R}\lesssim 0.4$, but fails to capture the rapid increase in the slope for $\gamma_{\rm R}\gtrsim 0.6$.


\subsection{Frequency-dependent electrical conductivity\label{sec:conduc}}
Here we investigate the frequency dependence of the electrical conductivity within the one-loop approximation. Throughout this sub-section, we use the following abbreviation,
\begin{align}
    \ulq=({\bm q},\omega),\quad \ulq'=(0,\omega').
\end{align}
Substituting Eqs.~(\ref{Je2}) and (\ref{dSdE}) into (\ref{cond2}), one easily finds that the following averages appear:
\begin{align}
    &\av{\phi_i(\ulq)V_\beta(-\ulq ')},\ \av{\phi_i(\ulq)W_{jk\beta}(-\ulq ')},\\
    &\av{Y_{i\alpha}(\ulq)V_\beta(-\ulq ')},\ \av{X_\alpha(\ulq)W_{ij\beta}(-\ulq ')},\\
    &\av{X_\alpha(\ulq)V_\beta(-\ulq ')},\ \av{Y_{i\alpha}(\ulq)W_{jk\beta}(-\ulq ')}. \label{nonvani}
\end{align}
At the one-loop level, however, one can easily see that the only nonvanishing averages are $\av{X_\alpha(\ulq)V_\beta(-\ulq')}$ and $\av{Y_{i\alpha}(\ulq) W_{jk\beta}(-\ulq')}$ in Eq.~(\ref{nonvani}). 

\subsubsection{Electrophoretic term\label{sec:electrophoretic}}
\begin{figure}[htp]
    \includegraphics[width=0.7\columnwidth]{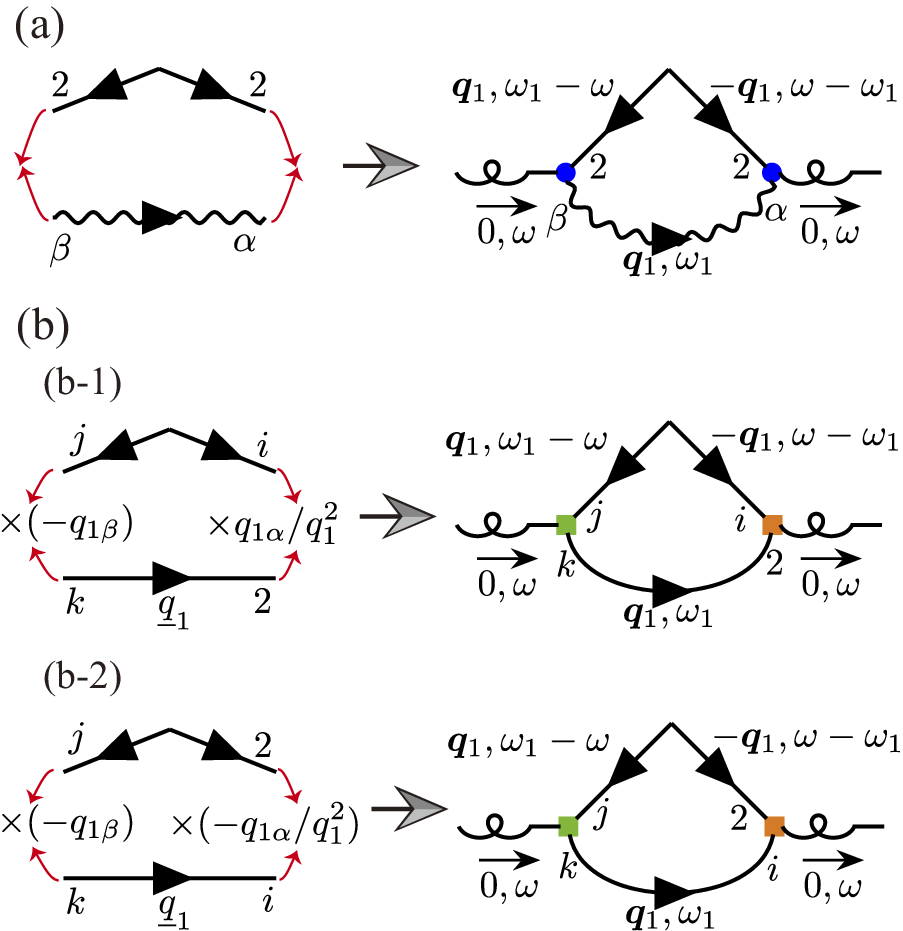}
    \caption{One-loop graphs of (a)$G_{XV}^{\alpha\beta}$ and (b)$G_{YW}^{ijk;\alpha\beta}$. Each graph in the right, which expresses $G_{XV}^{\alpha\beta}$ or $G_{YW}^{ijk;\alpha\beta}$, is made from the one in the left by contracting the sides of two propagators. \label{fig:sigma2}}
\end{figure}

First we calculate $\av{X_\alpha(\ulq)V_\beta(-\ulq')}$. Performing the Fourier transform of Eqs.~(\ref{def:X}) and (\ref{def:V}), we obtain
\begin{align}
    \av{X_\alpha(\ulq)V_\beta(-\ulq')} 
    =&\int_{\ulq_1,\ulq_2} \av{v_\alpha(\ulq_1)\phi_2(\ulq-\ulq_1)\tilde v_\beta(\ulq_2)\phi_2(-\ulq'-\ulq_2)} \\
    =&(2\pi)^4\delta_{\bm q}\delta_{\omega-\omega'}G_{XV}^{\alpha\beta}(\omega), \label{GXV}
\end{align}
where the delta functions in the last line appears due to the space-time translational invariance.
The four-point correlation function in the second line can be written as
\begin{align}
   \av{v_\alpha(\ulq_1)\phi_2(\ulq-\ulq_1)\tilde v_\beta(\ulq_2)\phi_2(-\ulq'-\ulq_2)} 
    =&\av{v_\alpha(\ulq_1)\phi_2(\ulq-\ulq_1)\tilde v_\beta(\ulq_2)\phi_2(-\ulq'-\ulq_2)}_{\rm c} \nn\\
    &+\av{v_\alpha(\ulq_1)\phi_2(\ulq-\ulq_1)}\av{\tilde v_\beta(\ulq_2)\phi_2(-\ulq'-\ulq_2)} \nn\\
    &+\av{v_\alpha(\ulq_1)\phi_2(-\ulq'-\ulq_2)}\av{\tilde v_\beta(\ulq_2)\phi_2(\ulq-\ulq_1)} \nn\\
    &+\av{v_\alpha(\ulq_1)\tilde v_\beta(\ulq_2)}\av{\phi_2(\ulq-\ulq_1)\phi_2(-\ulq'-\ulq_2)} \label{XV}
\end{align}
where $\av{\cdots}_{\rm c}$ denotes a connected correlation function (or cumulant), and we have used the fact that the first moment of any field vanishes. One can readily notice that only the last term of Eq.~(\ref{XV}) has the one-loop contribution to $G_{XV}^{\alpha\beta}(\omega)$. The corresponding graph is shown in Fig.~\ref{fig:sigma2}a; the contraction of external lines of $C_{\phi 0}^{22}G_{v0}^{\alpha\beta}$ in the left yields the one-loop graph of $G_{XV}^{\alpha\beta}(\omega)$ in the right. 
At zero frequency $\omega=0$, this contribution, involving the velocity propagator (wavy line), will turn out to be the one called the electrophoretic term\cite{RobinsonStokes}. Hence, even for non-zero frequencies, we shall call it the electrophoretic term. It is denoted by $\sigma_{\alpha\beta}^{\rm ph}$ and is expressed as
\begin{align}
    \sigma_{\alpha\beta}^{\rm ph}(\omega)=(e^2/\rho)\int_{\ulq_1}G_{v0}^{\alpha\beta}({\bm q}_1,\omega_1)C_{\phi 0}^{22}({\bm q}_1,\omega_1-\omega).
\end{align}
Substituting Eqs.~(\ref{Cp0}) and (\ref{Prop_v}) into the above, and performing $\omega_1$-integration, we obtain $\sigma_{\alpha\beta}^{\rm ph}=\sigma^{\rm ph}_{\rm B}\delta_{\alpha\beta}$ with the ``bare'' electrophoretic term $\sigma^{\rm ph}_{\rm B}$,
\begin{align}
    \sigma^{\rm ph}_{\rm B}(\omega)=&\frac{4\bar n e^2}{3\rho} \int_{q_1} \frac{q_1^2(i\omega+Dq_1^2+\nu_0q_1^2)}{(i \omega +\lambda_1+\nu_0q_1^2)(i\omega+\lambda_2+\nu_0q_1^2)(q_1^2+\kappa^2)}.
\end{align}
Here we have used the abbreviations, $\lambda_1=\lambda_1(q_1)$ and $\lambda_2=\lambda_2(q_1)$. Furthermore, Eq.~(\ref{schmidt}) enables us to neglect the $\gamma$-dependence of $\lambda_i$ in the sum $\lambda_i+\nu_0 q_1^2$. Hence, performing the angular-integration and changing the integration variable, $q_1 \to s=q_1/\kappa$, we obtain
\begin{align}
    \sigma^{\rm ph}_{\rm B}(\omega)=&\frac{2 \bar n e^2}{3\pi^2 D\rho} \int_0^{\Lambda/\kappa}  \hspace{-1mm}\frac{s^4\, ds}{(s^2+1)[i\Omega+1+\hat \nu_0 s^2]}\nn\\
    =& \frac{2 \kappa \bar n e^2}{3\pi^2 \eta_0} \Big[ (\Lambda/\kappa)+ \frac{\tan^{-1} (\Lambda/\kappa)}{a-1} -\frac{a^{3/2}}{a-1} \tan^{-1}(\Lambda/\kappa\sqrt{a})\Big], \label{sigma_bare}
\end{align}
where $a(\Omega;\hat\nu_0)=(i\Omega+1)\hat\nu_0^{-1}$. Here, our coarse-grained model cannot be valid for length-scales smaller than $\Lambda^{-1}$ and hence we must assume sufficiently small salt density such that $\Lambda/\kappa \gg 1$. We thus have $\tan^{-1}(\Lambda/\kappa)\approx \pi/2$, so
\begin{align}
    \sigma^{\rm ph}_{\rm B}(\omega)= \frac{2 \bar n e^2\Lambda}{3\pi^2 \eta_0} + \frac{\kappa\bar n e^2}{3\pi \eta_0 (a-1)}\Big[1-\frac{2a^{3/2}}{\pi}\tan^{-1}\frac{\Lambda}{\kappa\sqrt{a}}\Big]. \label{sigma:phoretic0}
\end{align}
The first term, which diverges in the limit $\Lambda\to\infty$, can be absorbed into the renormalized coefficient $D_{\rm R}$ by replacing $D$ by $D_{\rm R}$ in the Nernst-Einstein term $\sigma^{\rm NE}_{\rm B}$. That is, from Eqs.~(\ref{sigma_NE}) and (\ref{sigma:phoretic0}), we have $\sigma^{\rm NE}_{\rm B}+\sigma^{\rm ph}_{\rm B}=\sigma^{\rm NE}+\sigma^{\rm ph}$ with
\begin{align}
    &\sigma^{\rm NE}=2e^2D_{\rm R}\bar n/T=e^2(D_{\rm +R}+D_{\rm -R})\bar n/T \label{sigma:NE}\\
    &\sigma^{\rm ph}=\frac{\kappa\bar n e^2}{3\pi \eta_0 (a-1)}\Big[1-\frac{2a^{3/2}}{\pi}\tan^{-1}\frac{\Lambda}{\kappa\sqrt{a}}\Big].\label{sigma:phoretic01}
\end{align} 
Note that in the second line the dimensionless frequency $\Omega=\omega/D\kappa^2$ and the ratio $\hat \nu_0=\nu_0/D$ have implicitly been replaced by $\Omega=\omega/D_{\rm R}\kappa^2$ and $\hat \nu_0=\nu_0/D_{\rm R}$, respectively, in the spirit of perturbation theory.

The remaining $\Lambda$-dependent term, $\tan^{-1}(\Lambda/\kappa\sqrt{a})$, is insensitive to $\Lambda$ for sufficiently small frequencies; for $|\omega|\ll \nu_0\Lambda^2$, we have $|a|=\sqrt{\omega^2/(\nu_0^2\kappa^4)+1/\hat\nu_0^2}\ll \Lambda^2/\kappa^2$ and thus $\tan^{-1}(\Lambda/\kappa\sqrt{a})\simeq\pi/2$ yielding
\begin{align}
    \sigma^{\rm ph}(\omega)= -\frac{\kappa \bar n e^2}{3\pi \eta_0}\frac{1-a^{3/2}}{1-i\Omega/\hat\nu_0} \quad (|\omega| \ll \nu_0\Lambda^2). \label{sigma:phoretic}
\end{align}
Further, for $|\omega| \ll \nu_0\kappa^2 (\ll \nu_0\Lambda^2)$, we expand the above expression in powers of $\Omega/\hat \nu_0$ and obtain 
\begin{align}
    \sigma^{\rm ph}(\omega)\simeq -\frac{\kappa \bar n e^2}{3\pi \eta_0}\Big[1+\frac{i\Omega}{\hat\nu_0}+\frac{3\Omega^2}{8\hat\nu_0^{3/2}}\Big]. \label{sig_ph_app1}
\end{align}
 For $ \nu_0\kappa^2 \ll |\omega| (\ll \nu_0\Lambda^2)$, conversely, we have $a\simeq i\Omega/\nu_0$ and $|a|\gg 1$, so
\begin{align}
    \sigma^{\rm ph}(\omega)\simeq -\frac{\kappa \bar n e^2}{3\pi \eta_0 } \frac{|\Omega|^{1/2}}{\sqrt{2\hat\nu_0}} [1+i\,\mathrm{sgn}(\Omega)]. \label{sig_ph_app2}
\end{align}

For $|\omega|\gtrsim \nu_0\Lambda^2$, on the other hand, the last term of Eq.~(\ref{sigma:phoretic0}) is $\Lambda$-sensitive. This suggests that in this regime the frequency dependence is highly ion-specific and cannot be determined by the diffusion coefficients. The investigation of such ion-specific behavior is beyond the scope of the present theory, and we hereafter discuss the frequency dependence for frequencies such that
\begin{align}
    |\omega|\ll \nu_0\Lambda^2\quad \Leftrightarrow \quad |\Omega| \ll \hat\nu_0(\Lambda/\kappa)^2.
\end{align}
Note that one should not take the limit $\Lambda\to\infty$ in Eq.~(\ref{sigma:phoretic01}). In this limit, we have $\tan^{-1}(\Lambda/\kappa\sqrt{a})=\pi/2$ and hence Eq.~(\ref{sigma:phoretic01}) becomes the same as Eq.~(\ref{sigma:phoretic}) but in the \textit{entire range} of $\omega$; this is unphysical because the right hand side of Eq.~(\ref{sigma:phoretic}) diverges as $|\omega|\to \infty$. On the other hand, the ``bare'' electrophoretic part in Eq.~(\ref{sigma:phoretic0}) for finite $\Lambda$ vanishes in the high frequency limit, $|\omega|\to\infty$, as expected. 

\begin{figure}[htp]
    \includegraphics[width=0.8\columnwidth]{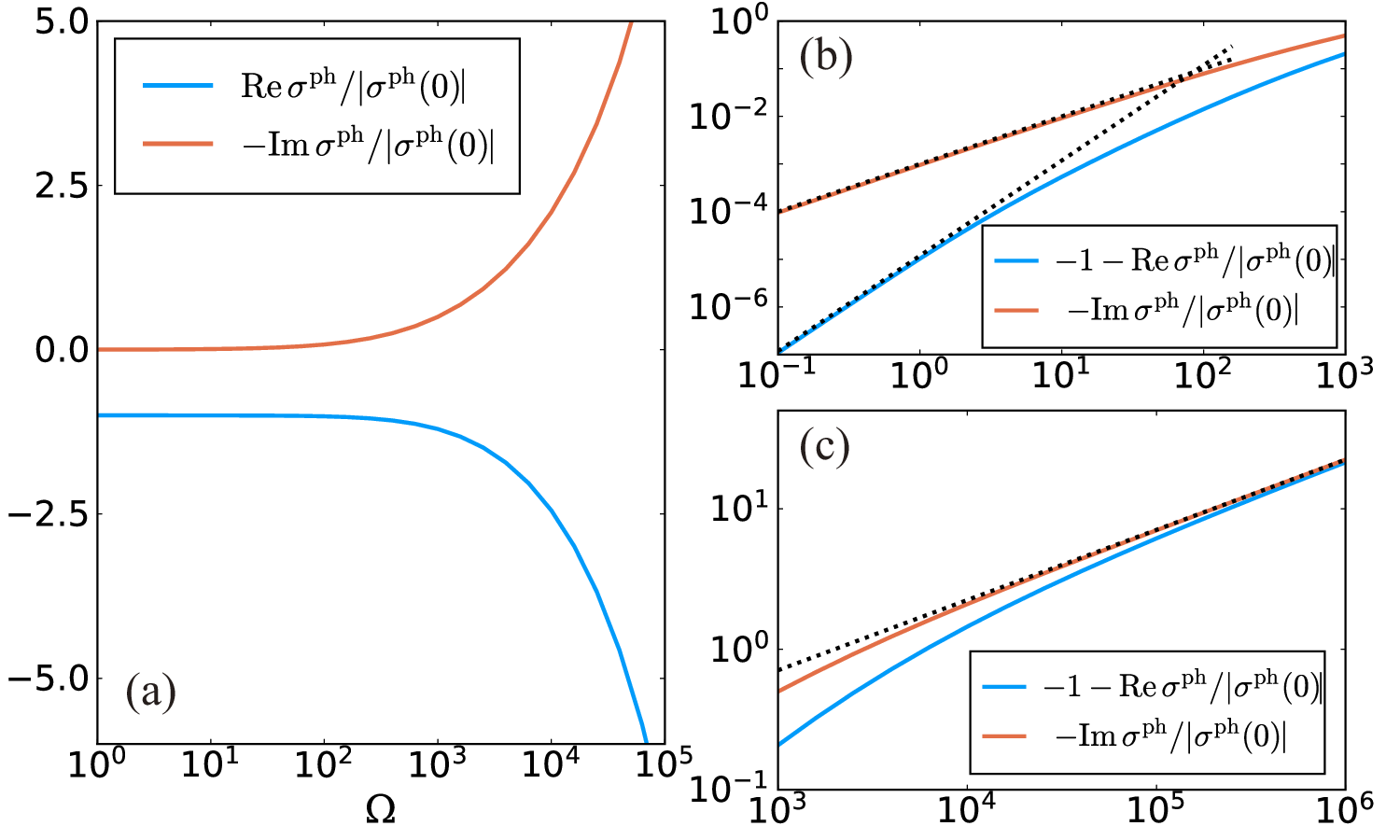}
    \caption{The electrophoretic term of the conductivity $\sigma^{\rm ph}(\omega)$ for low frequency range, $\Omega=\omega/D_{\rm R}\kappa^2 \ll \hat\nu_0(\Lambda/\kappa)^2$. (a) The real and minus the imaginary parts of $\sigma^{\rm rel}(\omega)/|\sigma^{\rm rel}(0)|$ with solid lines. (b) $-1-{\rm Re}\, \sigma^{\rm rel}/|\sigma^{\rm rel}(0)|$ and $-{\rm Im}\, \sigma^{\rm rel}/|\sigma^{\rm rel}(0)|$ for low frequency such that $\Omega \lesssim \hat\nu_0$ in log-log scales. The dotted lines show the approximate expression in Eq.~(\ref{sig_ph_app1}). (c) $-1-{\rm Re}\, \sigma^{\rm rel}/|\sigma^{\rm rel}(0)|$ and ${\rm Im}\, \sigma^{\rm rel}/|\sigma^{\rm rel}(0)|$ for higher frequency such that $\hat\nu_0\lesssim\Omega \ll\hat\nu_0(\Lambda/\kappa)^2$ in log-log scales. The dotted line shows the approximate expression in Eq.~(\ref{sig_ph_app2})\label{fig:sigPH}}
\end{figure}

Figure \ref{fig:sigPH}a shows the plots of Eq.~(\ref{sigma:phoretic}) for 
\begin{align}
    \hat \nu_0=\nu_0/D_{\rm R}=1000.0. \label{nu_value}
\end{align}
See also the sentence above Eq.~(\ref{schmidt}). The both lines of ${\rm Re}\, \sigma^{\rm ph}$ and $-{\rm Im}\, \sigma^{\rm ph}$ are almost flat up to $\Omega\sim \hat\nu_0$, i.e., $\omega \sim \nu_0\kappa^2$. This can be understood from the approximate expression Eq.~(\ref{sig_ph_app1}) where the $\Omega$-dependent term is small for $\Omega\lesssim \hat\nu_0$; Figure \ref{fig:sigPH}b shows a good agreement between the approximate expression (dotted lines) and $\sigma^{\rm ph}(\omega)$. As $\Omega$ is further increased so that $\hat\nu_0\lesssim \Omega $, the real part ${\rm Re}\, \sigma^{\rm ph}$ decreases and hence the electrophoretic effect tends to suppress the conductivity as $|\omega|$ is increased. In contrast, as is well-known, the relaxation effect tends to enhance the real part of the conductivity for high frequencies, which is referred to as the Debye-Falkenhagen effect (see Sec.~\ref{sec:relax}). In accordance with ${\rm Re}\, \sigma^{\rm ph}$, the imaginary part $-{\rm Im}\, \sigma^{\rm ph}$ also grows for $\hat\nu_0\lesssim\Omega$ as is shown in Fig.~\ref{fig:sigPH}a. For $\hat\nu_0 \ll \Omega\ll \hat\nu_0(\Lambda/\kappa)^2$, the growths in ${\rm Re}\, \sigma^{\rm ph}$ and $-{\rm Im}\, \sigma^{\rm ph}$ are well expressed by Eq.~(\ref{sig_ph_app2}) which is plotted with a dotted line in Fig.~\ref{fig:sigPH}c.

\subsubsection{Relaxation term \label{sec:relax}}
We now discuss the relaxation contribution that arises from the one-loop approximation of $\av{Y_{i\alpha}(\ulq) W_{jk\beta}(-\ulq')}$. Owing to the space-time translational invariance, $\av{Y_{i\alpha}(\ulq) W_{jk\beta}(-\ulq')}$ is written in the form
\begin{align}
    \av{Y_{i\alpha}(\ulq) W_{jk\beta}(-\ulq')}=(2\pi)^4\delta_{\bm q}\delta_{\omega-\omega'}G_{YW}^{ijk;\alpha\beta}(\omega)
\end{align}
as in Eq.~(\ref{GXV}).
This is a four-body correlation function of $\phi$ and $\tilde\phi$ (with two pairs of the external lines being contracted), so it can be decomposed into one connected four-body correlation function and three products of two-body functions as in Eq.~(\ref{XV}); at the one-loop level, two of the three products are nonvanishing and are expressed by the graphs in Fig.~\ref{fig:sigma2}b. The corresponding analytical expression is 
\begin{align}
    &G_{YW}^{ijk;\alpha\beta}(\omega)=4\pi\ell_{\rm B} \int_{\ulq_1} \hat q_{1\alpha} \hat q_{1\beta} \big[ G_{\phi 0}^{2k}({\bm q}_1,\omega_1)C_{\phi 0}^{ij}({\bm q}_1,\omega_1-\omega) -G_{\phi 0}^{ik}({\bm q}_1,\omega_1)C_{\phi 0}^{2j}({\bm q}_1,\omega_1-\omega) \big], \label{GYW}
\end{align}
where the first and the second terms in the bracket are from the graphs (b-1) and (b-2), respectively.
It will turn out that the contribution from $G_{YW}^{ijk;\alpha\beta}$ is the one referred to as the relaxation term, originally derived by Debye and Falkenhagen\cite{DebyeFalken}. As was pointed out by Chandra and Bagchi\cite{ChandraBagchi2000,ChandraBagchi2000_2}, the expression of Debye and Falkenhagen is not the complete limiting law for the frequency dependence because the electrophoretic effect that has been discussed in Sec.~\ref{sec:electrophoretic} is not taken into account.

The relaxation contribution, denoted by $\sigma_{\alpha\beta}^{\rm rel}(\omega)$, is expressed in terms of $G_{YW}^{ijk;\alpha\beta}(\omega)$ as
\begin{align}
    \sigma_{\alpha\beta}^{\rm rel}=-\frac{D^2e^2}{T}\Big[&\gamma^2\{G_{YW}^{211;\alpha\beta}+G_{YW}^{222;\alpha\beta}\} +\gamma\{G_{YW}^{111;\alpha\beta} +G_{YW}^{122;\alpha\beta}+G_{YW}^{212;\alpha\beta}+G_{YW}^{221;\alpha\beta}\} \nn \\ 
    &+G_{YW}^{112;\alpha\beta}+G_{YW}^{121;\alpha\beta} 
    \Big].\label{sigma_rel}
\end{align}
Substituting Eq.~(\ref{GYW}) into (\ref{sigma_rel}) and using (\ref{Gp0}), we obtain $\sigma_{\alpha\beta}^{\rm rel}=\sigma^{\rm rel}\delta_{\alpha\beta}$ with
\begin{align}
    \sigma^{\rm rel}(\omega) 
    =\frac{4\pi\ell_{\rm B}D^2e^2}{3T}\int_{\ulq_1} \frac{1}{(i\omega_1+\lambda_1)(i\omega_1+\lambda_2)} \big[ &(i\omega_1+D(1-\gamma^2)(q_1^2+\kappa^2))C_{\phi 0}^{22}(q_1,\omega_1-\omega) \nn\\
    &\hspace{-5mm} -(i\omega_1+D(1-\gamma^2)q_1^2)C_{\phi 0}^{11}(q_1,\omega_1-\omega)\big].
\end{align}
Substituting Eq.~(\ref{Cp0}) and performing the integration, we obtain the classical result\cite{DebyeFalken}
\begin{align}
    \sigma^{\rm rel}(\omega)=-  \frac{\ell_{\rm B} D_{\rm R}e^2\bar n\kappa}{3 T} \frac{\sqrt{2}}{\sqrt{2}+\sqrt{1+i\Omega}}
    \label{sigma_rel2}
\end{align} 
where use has been made of Eq.~(\ref{eigen}), and we have again made the replacements $D \to D_{\rm R}$ and $\Omega=\omega/D\kappa^2\to \Omega=\omega/D_{\rm R}\kappa^2$ in the logic of perturbation theory. Notice that Eq.~(\ref{sigma_rel2}) is independent of the asymmetric factor $\gamma_{\rm R}$. 

Although Eq.~(\ref{sigma_rel2}) is thoroughly studied in the literature\cite{DebyeFalken}, in the following we shall see its overall behavior to investigate later the total conductivity including the electrophoretic term. For small frequencies $|\omega| \ll D_{\rm R}\kappa^2$ ($|\Omega|\ll 1$), we expand Eq.~(\ref{sigma_rel2}) with respect to $\Omega$ to have
\begin{align}
    \sigma^{\rm rel}(\omega) \simeq  \frac{\ell_{\rm B} D_{\rm R}e^2\bar n\kappa}{3 T}\Bigg[\frac{-\sqrt{2}}{1+\sqrt{2}}+iA\Omega +B\Omega^2 \Bigg], \label{sigDFsmall}
\end{align}
where $A=(3/\sqrt{2})-2\approx 0.12$ and $B=2-(11\sqrt{2}/8)\approx 0.055$ are positive numbers. 

For $|\omega| \gg D_{\rm R}\kappa^2$ ($|\Omega|\gg 1$), Eq.~(\ref{sigma_rel2}) yields
\begin{align}
    \sigma^{\rm rel}(\omega)\simeq - \frac{\ell_{\rm B} D_{\rm R}e^2\bar n\kappa}{3 T}f_{\rm rel}(\Omega) \label{sigDFlarge1}
\end{align}
with
\begin{align}
    f_{\rm rel}(x)=&\frac{2}{|x|+(|x|^{1/2}+2)^2} \big[(|x|^{1/2}+2) -i|x|^{1/2}\mathrm{sgn}(x) \big] \nn\\
    \simeq & |x|^{-1/2} -i |x|^{-1/2}\mathrm{sgn}(x) \quad (|x|^{1/2}\gg 1). \label{sigDFlarge2}
\end{align}
Hence for $|\omega|\gg D_{\rm R}\kappa^2$, the electrophoretic part $\sigma^{\rm ph}$ dominates over the relaxation part $\sigma^{\rm rel}$.

\begin{figure}[htp]
    \includegraphics[width=0.8\columnwidth]{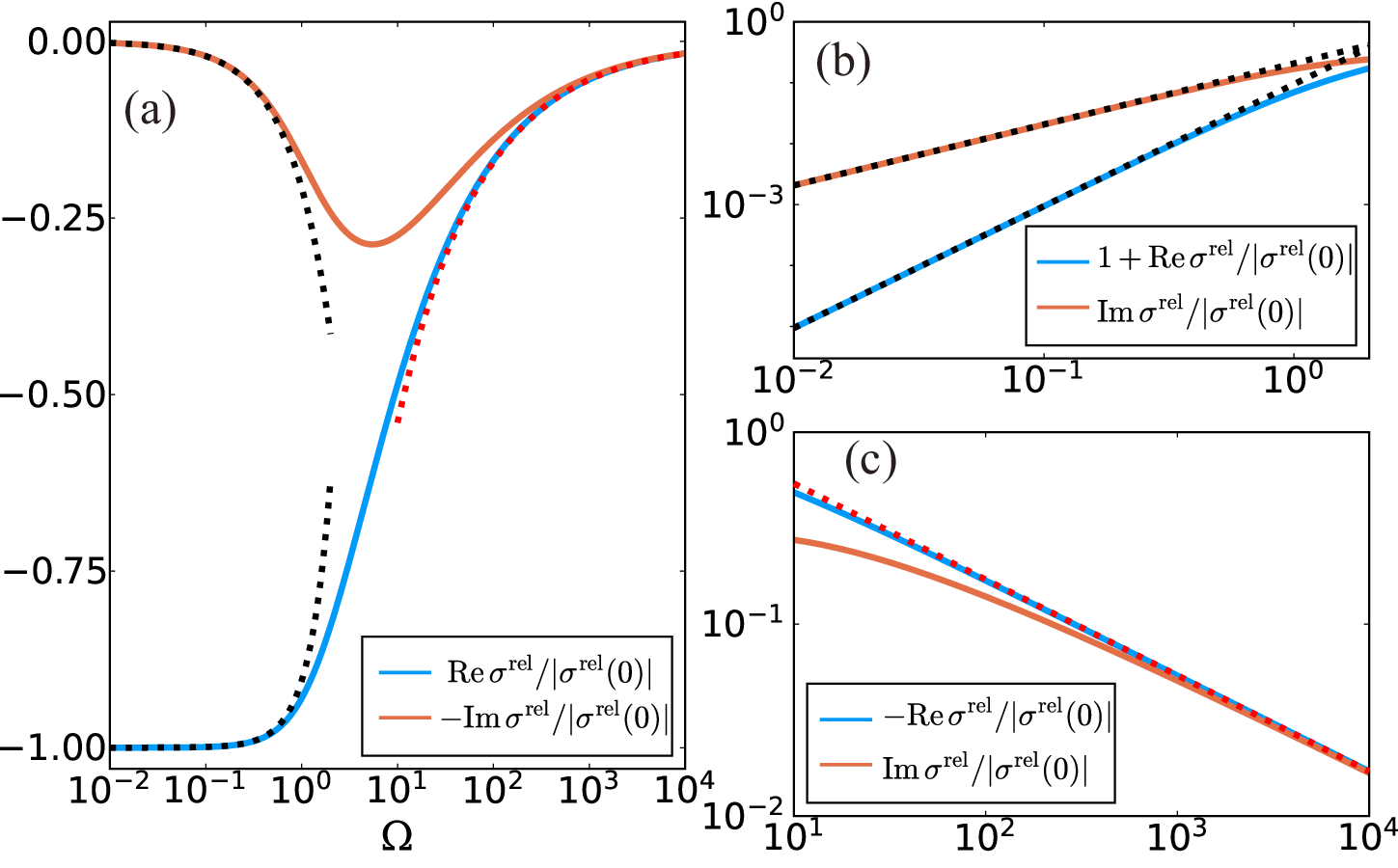}
    \caption{The relaxation term of the conductivity $\sigma^{\rm rel}(\omega)$ as a function of the dimensionless frequency $\Omega=\omega/D_{\rm R}\kappa^2$. (a) The real and minus the imaginary parts of $\sigma^{\rm rel}(\omega)/|\sigma^{\rm rel}(0)|$ with solid lines and the approximate expressions in Eqs.~(\ref{sigDFsmall}), (\ref{sigDFlarge1}), and (\ref{sigDFlarge2}) with dotted lines. (b) $1+{\rm Re}\, \sigma^{\rm rel}/|\sigma^{\rm rel}(0)|$ and ${\rm Im}\, \sigma^{\rm rel}/|\sigma^{\rm rel}(0)|$ for small $\Omega$ in log-log scales. (c) $-{\rm Re}\, \sigma^{\rm rel}/|\sigma^{\rm rel}(0)|$ and ${\rm Im}\, \sigma^{\rm rel}/|\sigma^{\rm rel}(0)|$ for large $\Omega$ in log-log scales.\label{fig:sigDF}}
\end{figure}

Figure \ref{fig:sigDF}a shows the real and imaginary parts of the relaxation term $\sigma^{\rm rel}/|\sigma^{\rm rel}(0)|$ as functions of $\Omega=\omega/D_{\rm R}\kappa^2$ calculated from Eq.~(\ref{sigma_rel2}), where $\sigma^{\rm rel}(0)=-\sqrt{2}\ell_{\rm B}D_{\rm R}e^2\bar n\kappa/3T(\sqrt{2}+1)$ is the zero-frequency relaxation term. We can see that the real part is an increasing function of $|\omega|$ varying from -1 to 0, which indicates that, (at lease) in the absence of the electrophoretic term, the real part of the conductivity increases with increasing frequency (Debye-Falkenhagen effect). We shall see later that this effect remains up to the maximum frequency $\omega_{\rm max}$ even in the presence of the electrophoretic term. On the other hand, the imaginary part ${\rm Im}\, \sigma^{\rm rel}$ has an extremum at around $\Omega \sim 1$ where the real part ${\rm Re}\, \sigma^{\rm rel}$ exhibits a steep increase.

In Fig.~\ref{fig:sigDF}a, the approximate expression for a small $|\Omega|$-range in Eq.~(\ref{sigDFsmall}) and that for a large $|\Omega|$-range in Eqs.~(\ref{sigDFlarge1}) and (\ref{sigDFlarge2}) are also plotted with black dotted lines and a red dotted line, respectively. In Fig.~\ref{fig:sigDF}b, similar plots are shown for a small $|\Omega|$-range in log-log scales; we can see the approximate expression in Eq.~(\ref{sigDFsmall}) is in good agreement with Eq.~(\ref{sigma_rel2}) for $\Omega\lesssim 1$. Similarly, Fig.~\ref{fig:sigDF}c shows that Eq.~(\ref{sigDFlarge1}) with Eq.~(\ref{sigDFlarge2}) is a good approximation in the high frequency range, $\Omega\gg 1$.

\subsubsection{Total conductivity and its maximum frequency}
Collecting Eqs.~(\ref{sigma:NE}), (\ref{sigma:phoretic}), and (\ref{sigma_rel2}) we obtain the total conductivity $\sigma=\sigma^{\rm NE}+\sigma^{\rm ph}+\sigma^{\rm rel}$,
\begin{align}
    \sigma(\omega)=&\frac{2e^2D_{\rm R}\bar n}{T}-\frac{\kappa \bar n e^2}{3\pi \eta_0}\frac{1-a^{3/2}}{1-i\Omega/\hat\nu_0} -  \frac{\ell_{\rm B} D_{\rm R}e^2\bar n\kappa}{3 T} \frac{\sqrt{2}}{\sqrt{2}+\sqrt{1+i\Omega}}, \label{sigma_tot}
\end{align}
where $a=a(\Omega;\hat\nu_0)$ has been defined below Eq.~(\ref{sigma_bare}).
Again, note that this expression, precisely the electrophoretic term, is valid for $\Omega\ll\hat \nu_0(\Lambda/\kappa)^2$. At $\Omega=0$, this reduces to the Onsager's limiting law\cite{Onsager1927,RobinsonStokes} for the zero-frequency conductivity,
\begin{align}
    \sigma(0)=&\frac{2e^2D_{\rm R}\bar n}{T}-\frac{\kappa \bar n e^2}{3\pi \eta_0} -  \frac{\ell_{\rm B} D_{\rm R}e^2\bar n\kappa}{3 T} \frac{\sqrt{2}}{\sqrt{2}+1}.
\end{align}
Recently, the above zero-frequency conductivity has been extended to more concentrated solutions by taking into account the hard-sphere-like repulsion between ions in FHE \cite{Avni2022}.

\begin{figure}[htp]
    \includegraphics[width=0.8\columnwidth]{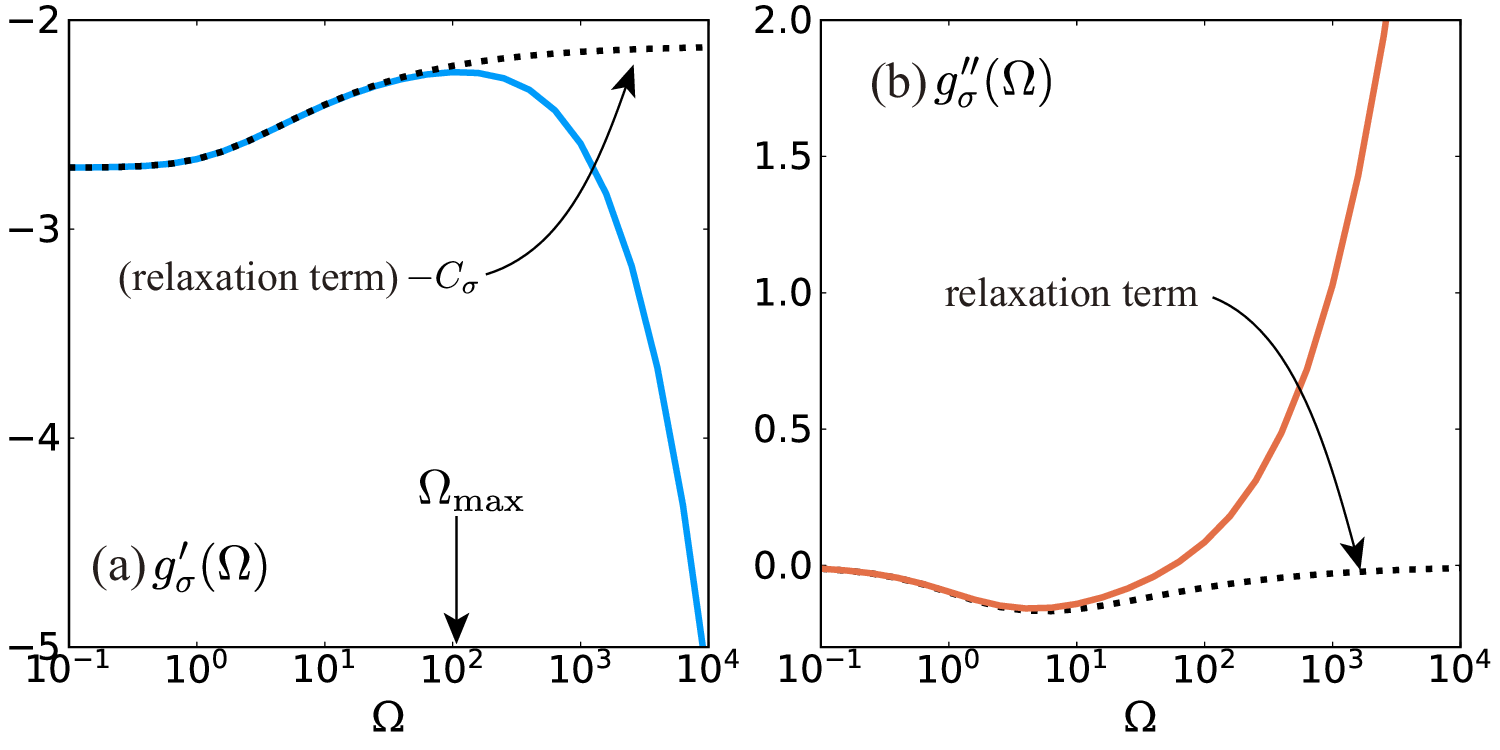}
    \caption{Scaled total conductivity as a function of $\Omega=\omega/D_{\rm R}\kappa^2$. (a) The shifted, scaled real part $g'_\sigma$ defined in Eq.~(\ref{gp}). (b) The scaled imaginary part $g''_\sigma$ defined in Eq.~(\ref{gpp}). In each panel, the dotted line shows the relaxation term [the second term of Eq.~(\ref{g_def})].\label{fig:sigTOT}}
\end{figure}

In the following, we shall discuss the frequency dependence of the conductivity normalized by the Nernst-Einstein term, i.e., $\sigma(\omega)/\sigma^{\rm NE}$. In this combination, besides $\hat\nu_0$ and $\Omega$, there are two dimensionless parameters, $\ell_{\rm B}\kappa$ and $C_\sigma\equiv T/(\pi\eta_0D_{\rm R}\ell_{\rm B})$. For the typical parameter values $\eta_0=0.89\times 10^{-3}\,{\rm Pa}\cdot {\rm s}$, $D_{\rm R}=10^{-9}\, {\rm m}^2/{\rm s}$, $\ell_{\rm B}=7$\AA, and $T=300\times 1.38\times 10^{-23}\,{\rm J}$, we have  
\begin{align}
    C_\sigma=2.12.\label{para}
\end{align} 
From Eq.~(\ref{sigma_tot}), we readily obtain
\begin{align}
    \frac{\sigma(\omega)}{\sigma^{\rm NE}}=1+\frac{\ell_{\rm B}\kappa}{6} g_\sigma(\Omega;\hat\nu_0,C_\sigma), \label{sig_norm}
\end{align}
where
\begin{align}
    g_\sigma=-C_\sigma \frac{1-a^{3/2}}{1-i\Omega/\hat\nu_0}-\frac{\sqrt{2}}{\sqrt{2}+\sqrt{1+i\Omega}}. \label{g_def}
\end{align}
Here, the first and second terms respectively correspond to the electrophoretic and relaxation effects.
As in the literature, we define $\sigma'={\rm Re}\, \sigma$ and $\sigma''=-{\rm Im}\, \sigma$, so
\begin{align}
    \sigma(\omega)=\sigma'(\omega)-i\sigma''(\omega).
\end{align}
Accordingly, we define $g_\sigma'={\rm Re}\, g_\sigma$ and $g_\sigma''=-{\rm Im}\, g_\sigma$. Then Eq.~(\ref{sig_norm}) yields
\begin{align}
    &\sigma'(\omega)/\sigma^{\rm NE}=1+(\ell_{\rm B}\kappa/6) g_\sigma' \label{gp}\\
    &\sigma''(\omega)/\sigma^{\rm NE}=(\ell_{\rm B}\kappa/6) g_\sigma''.\label{gpp}
\end{align}
Figure \ref{fig:sigTOT}a presents the plot of real part $g'_\sigma$ as a function of $\Omega$, where the parameters are set as in Eqs.~(\ref{nu_value}) and (\ref{para}). The shifted relaxation term $-C_\sigma-{\rm Re}\, [\sqrt{2}/(\sqrt{2}+\sqrt{1+i\Omega})]$ is also plotted with a dotted line. For $\Omega \lesssim 10^2$, $g'_\sigma$ increases with increasing $\Omega$ and the frequency dependence is dominated by the relaxation term. That is, the Debye-Falkenhagen (DF) effect remains even in the presence of the electrophoretic effect, of which frequency dependence is negligibly small for $|\Omega|\ll \hat\nu_0$ [See Eq.~(\ref{sig_ph_app1})]. In fact, Debye-Falkenhagen theory of $\sigma^{\rm rel}$ followed the Wien's experiment that suggests the increase in $\sigma'$ with increasing $\omega$ \cite{Anderson1994,Wien1927,Wien1928}.
As $\Omega$ further increases, $g'_\sigma$ exhibits a maximum at $\Omega=\Omega_{\rm max}\sim 10^2$ and for $\Omega_{\rm max} \ll |\Omega| (\ll \hat \nu_0\Lambda^2/\kappa^2)$, the conductivity is suppressed by the electrophoretic effect as $g'_\sigma \sim -|\Omega|^{1/2}$ [See Eq.~(\ref{sig_ph_app2})]. 

Let us here discuss in detail the maximum frequency $\omega_{\rm max}= \Omega_{\rm max}D_{\rm R}\kappa^2$ at which $\sigma'$ has a maximum. Previously, a molecular dynamics simulation\cite{Chandra1993} and numerical analysis based on the MCT\cite{ChandraBagchi2000,ChandraBagchi2000_2} also showed that the real part of conductivity exhibits similar non-monotonic behavior. 
Because $\Omega_{\rm max}$ is determined by $\p g'_\sigma(\Omega;\hat \nu_0,C_\sigma)/\p \Omega=0$, it is a function of $\hat\nu_0$ and $C_\sigma$, which are independent of the salt density or $\kappa$. Hence we have
\begin{align}
    \omega_{\rm max}=D_{\rm R}\kappa^2 \Omega_{\rm max}(\hat\nu_0,C_\sigma)\propto \bar n.
\end{align}
That is, the frequency $\omega_{\rm max}$ is proportional to the salt density $\bar n$. This is in semi-quantitative agreement with the previous numerical results based on the MCT\cite{ChandraBagchi2000,ChandraBagchi2000_2}. Meanwhile, to the best of the author's knowledge, there is no experimental data of $\omega_{\rm max}$ in the literature.

As shown in Appendix \ref{sec:omega_max}, the reduced maximum frequency $\Omega_{\rm max}$ is approximately given by
\begin{align}
    \Omega_{\rm max}\approx\hat\nu_0 \left[ \Big(\frac{\epsilon_\sigma}{3}\Big)^{1/4}+\frac{\sqrt{6\epsilon_\sigma}}{9}\right]^2,\label{Omega_max_app}
\end{align}
where we have introduced a small parameter $\epsilon_\sigma=\sqrt{2}/(\hat\nu_0^{1/2}C_\sigma)$. For the typical parameter values in Eqs.~(\ref{nu_value}) and (\ref{para}), we have $\epsilon_\sigma=0.021$, and hence the right hand side of Eq.~(\ref{Omega_max_app}) yields $\Omega_{\rm max}\approx 108.2$, which is in good agreement with the numerical result $\Omega_{\rm max}=112.7$. For different parameter values, $C_\sigma=1.0$ (resp.~$3.0$), we obtain $\Omega_{\rm max}\approx 165.6$ (resp.~$89.2$) from Eq.~(\ref{Omega_max_app}), while the numerically obtained values are $\Omega_{\rm max}=177.6$ (resp.~$91.8$) [Here, the value of $\hat\nu_0$ is fixed as in Eq.~(\ref{nu_value})]. As a concrete example, let us examine an aqueous NaCl solution in ambient conditions. Using the parameter values $D_{\rm +R}=1.3\times 10^{-9}$m$^2$/s,, $D_{\rm -R}=2.0\times 10^{-9}$m$^2$/s,  $\eta_0=0.89\times 10^{-3}\,{\rm Pa}\cdot {\rm s}$, and $\ell_{\rm B}=7$\AA, we obtain $\hat\nu_0=545.5$ and $C_\sigma=1.28$ yielding $\Omega_{\rm max}=99.2$. Further, at the salt concentration of 0.03 molar, for instance, the Debye wavenumber is $\kappa=0.32$ nm$^{-1}$, and we hence have $\omega_{\rm max}=D_{\rm R}\kappa^2\Omega_{\rm max}=1.7\times 10^8$ s$^{-1}$.

Figure \ref{fig:sigTOT}b presents the plot of the imaginary part $g''_\sigma$ with a solid line and the relaxation term $-{\rm Im}\, [\sqrt{2}/(\sqrt{2}+\sqrt{1+i\Omega})]$ with a dotted line. As in the case of $g'_\sigma$, it is dominated by the relaxation term for small $\Omega \ll \hat\nu_0$, and has a minimum which we have also seen in Fig.~\ref{fig:sigDF}b for the relaxation term. This is coherent with the Debye-Falkenhagen effect in the real part that remains even with the electrophoretic term. The decrease in $\sigma''$ with increasing $\omega$ is equivalent to the increase in the complex dielectric permittivity with increasing $\omega$, which has been detected in the previous experiment\cite{Hubbard1977}. For $\hat\nu_0\lesssim \Omega \ll \hat\nu_0(\Lambda/\kappa)^2$, the electrophoretic term becomes the dominant and $g''_\sigma$ grows as $g''_\sigma\sim \Omega^{1/2}$ [See Eq.~(\ref{sig_ph_app2})].

\section{Discussion\label{sec:discussion}}
In the previous sections, we have discussed the dynamic properties of electrolyte solutions, in particular, the frequency-dependent viscosity and conductivity, by performing systematic perturbation calculations. At zero frequency, the viscosity and the conductivity reduce to the classical results by Debye, Falkenhagen, Onsager, and others\cite{falkenhagenLXIIViscosityStrong1932,Onsager1927,RobinsonStokes}.
These results have been obtained within the one-loop approximation. We have also assumed the local free energy density in Eq.~(\ref{dilute}). 


One may then ask what we expect if we go beyond the one-loop level and/or the free energy density of Eq.~(\ref{dilute}) [See also the sentences below Eq.~(\ref{dilute})]. The full investigation is beyond the scope of the present study, but we shall discuss how these higher-order effects contribute, as an example, to the effective viscosity $\eta_{\rm eff}$ in slightly more concentrated solutions.

The (zero-frequency) effective viscosity $\eta_{\rm eff}$ is known to obey the Jones-Dole empirical expression\cite{jonesVISCOSITYAQUEOUSSOLUTIONS1929}, 
\begin{align}
    \eta_{\rm eff}/\eta_0=1+A\bar n^{1/2}+B\bar n+\cdots.
\end{align}
The first correction $A\bar n^{1/2}\propto \kappa$ is indeed given by the Falkenhagen's expression\cite{falkenhagenLXIIViscosityStrong1932}, or, equivalently, the one-loop result in Eq.~(\ref{vis_eff_0}). We can easily ``anticipate'' that the one-loop result is proportional to $\kappa$ without performing the explicit calculation. In the analytical expression
Eq.~(\ref{self_v}) of the graph in Fig.~\ref{fig:graph_vis}, let us make all the frequency- and the wavenumber-variables dimensionless, e.g., ${\bm q}\to {\bm q}/\kappa$ and $\omega_1 \to \omega_1/D\kappa^2$. Then we count the exponent of $\kappa$ in Eq.~(\ref{self_v}). First we consider the two density correlation propagators $C_{\phi 0}$. We readily see that $C_{\phi 0}({\bm q},\omega)$ in Eq.~(\ref{Cp0}) can be written in the form, 
\begin{align}
    C_{\phi 0}({\bm q},\omega)=\bar n(D\kappa^2)^{-1}  g_{\phi0}({\bm q}/\kappa, \omega/D\kappa^2;\gamma).
\end{align}
Hence, noting that $\bar n\propto \kappa^2$, we see that the exponent of $\kappa$ is zero for $C_{\phi 0}$. Similarly, for each of two vertex factors $w_\alpha$ and $w_\beta$, the exponent is $-1$. Finally, the integral $\int_{\ulq_1}$ has $5$. Therefore the graph in Fig.~\ref{fig:graph_vis} can be written in the form,
\begin{align}
    \mbox{Fig.~\ref{fig:graph_vis}}=\kappa^{0+0-1-1+5} g_{v,{\rm 1loop}}({\bm q}/\kappa, \omega/D\kappa^2), \label{diss_oneloop}
\end{align}
where in $g_{v,{\rm 1loop}}$ the dependence on the other parameters such as $D$ has been omitted because they are irrelevant to the present purpose. The one-loop correction to the zero-frequency viscosity is obtained by differentiating Eq.~(\ref{diss_oneloop}) with respect to $q^2$ and setting $q=\omega=0$. Therefore we can easily see that the correction is indeed proportional to $\kappa$. 

\begin{figure}[htp]
    \includegraphics[width=0.7\columnwidth]{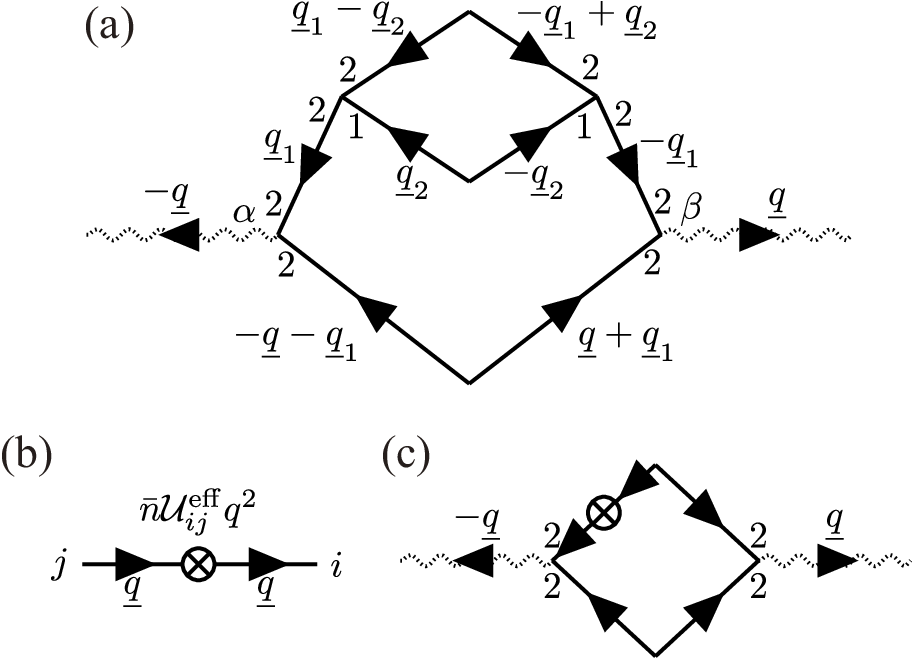}
    \caption{(a) An example of a two-loop graph for the effective viscosity. (b) Bilinear term in the action due to the short-range interaction between ions. (c) An example of a one-loop graph with a correction by the short-range interaction. \label{fig:graph_vis2}}
\end{figure}

We can apply the same argument to the two-loop case. We here assume the free energy density of Eq.~(\ref{dilute}). Figure \ref{fig:graph_vis2}a shows an example of a two-loop graph for $\Sigma_{v,\alpha\beta}^{(0,2)}$. It has three $C_{\phi 0}$'s and two $G_{\phi 0}$'s, where the former has the exponent $0$ while each $G_{\phi 0}$ has $-2$. Futhermore, each of the two vertex factors $v_{1;2}$ has the exponent $0$. As in the one-loop case, two vertex factors $w_\alpha$ and $w_\beta$ have $-2$. Finally, the integral $\int_{\ulq_1}\int_{\ulq_2}$ has $10$. We thus obtain
\begin{align}
    \mbox{Fig.~\ref{fig:graph_vis2}a}=\kappa^{4} g_{v,{\rm 2loop}}({\bm q}/\kappa, \omega/D\kappa^2), \label{diss_twoloop}
\end{align}
yielding the two-loop correction to the viscosity proportional to $\kappa^2\propto \bar n$. That is, the two-loop graph in Fig.~\ref{fig:graph_vis2}a contributes to the $B$-coefficient in the Jones-Dole expression.

Finally, we examine the effect of inter-ionic short-range interactions, which we take into account by adding the following term in the local free energy\cite{Okamoto2018},
\begin{align}
    f_{\rm int}=\frac{1}{2}\sum_{i,j=+,-}U_{ij}^{\rm eff}n_in_j. \label{f_int}
\end{align}
Here, the sum of the coefficients $U^{\rm eff}_{++}+U^{\rm eff}_{--}+2U^{\rm eff}_{+-}$ is experimentally accessible by a measurement of the salt activity coefficient\cite{Okamoto2020}. We can readily see that $f_{\rm int}$ gives rise to not only new three-point vertices in the action but also a modification of the propagators in Eqs.~(\ref{Gp0}) and (\ref{Cp0}). Because the new vertices have no velocity fields, there is no one-loop graph of $\Sigma_{v}^{(0,2)}$ that contains the new vertices. We here discuss only the effect of the modified propagators at the one-loop level. With the interaction term in Eq.~(\ref{f_int}), besides three-point vertex terms, there appears new terms in the bilinear part of the form,
\begin{align}
    S_0^{\rm int}=\bar n\int_{\ulq} q^2\sum_{i,j=1,2} {\cal U}_{ij}^{\rm eff}\tilde\phi_i\phi_j, \label{Sint}
\end{align}
where ${\cal U}_{ij}^{\rm eff}$ $(i,j=1,2)$ is given by a linear combination of $U_{ij}^{\rm eff}$'s in Eq.~(\ref{f_int}). This new bilinear terms modify the propagators, but we here regard $S_0^{\rm int}$ as a perturbation instead of using the full modified propagators. That is, we perform perturbation expansion with respect to the number of loops \textit{and} the bilinear interaction term $S_0^{\rm int}$. Figures \ref{fig:graph_vis2}b and c respectively present the graph representation of the bilinear term in Eq.~(\ref{Sint}) and an example of a one-loop graph containing the interaction term ${\cal U}_{ij}^{\rm eff}$ as a perturbation. We again count the exponent of $\kappa$ of the latter graph as in the previous case and readily find the form, 
\begin{align}
    \mbox{Fig.~\ref{fig:graph_vis2}c}=\kappa^5 g_{v,{\rm 1loop}}^{\rm int}({\bm q}/\kappa,\omega/D\kappa^2).
\end{align}
Therefore this graph gives a correction proportional to $\kappa^{3}\propto \bar n^{3/2}$, and thus no contribution to the Jones-Dole $B$-coefficient. Previous experimental studies have shown that the $B$-coefficient can empirically been split into the additive contributions from individual ion species, suggesting that the inter-ionic short-range interactions are irrelevant to the $B$-coefficient\cite{CoxWolfenden1934,JenkinsMarcus1995}.

\section{Summary\label{sec:summary}}
In this paper, we have presented a systematic perturbation theory of fluctuating hydrodynamics of electrolyte solutions. Our starting point is the coupled nonlinear Langevin equations in Eqs.~(\ref{diffusion}) and (\ref{deqv}), in the former of which the noise term enters as a multiplicative one. Under It\^o prescription, we have transformed the set of equations into the MSRJD path-integral form in Sec.~\ref{sec:MSRJD}. In Sec.~\ref{sec:vertex}, we have seen the effective kinetic coefficients are naturally written in terms of the associated vertex functions. In Sec.~\ref{sec:conduc0}, the frequency-dependent electrical conductivity has also been expressed as a sum of correlation functions of the physical and auxiliary fields.

In Sec.~\ref{sec:result} we have studied the dynamic properties within the one-loop approximation. In Sec.~\ref{sec:Lij}, the effective kinetic coefficients for the density fields have been studied in the limit of small wavenumber and frequency. We have also introduced the renormalized diffusion coefficients into which the cutoff sensitivity is absorbed\cite{Peraud2017,Donev2019}. In particular, our straightforward calculation derived the cross diffusion coefficient $L_{+-}^{\rm eff}$ between cations and anions, while the cross coefficient is absent in the starting Langevin equation in Eq.~(\ref{diffusion}). This has previously been anticipated from the zero-frequency conductivity for symmetric salts\cite{Peraud2017}.

In Sec.~\ref{sec:vis} we have investigated the frequency dependence of the effective shear viscosity $\eta_{\rm eff}(\omega)$. At the one-loop level, we have seen that $D_{\pm \rm R}\kappa^2$ are the only characteristic frequencies in $\eta_{\rm eff}(\omega)$. That is, the frequency dependence of $\eta_{\rm eff}(\omega)$ is dominated by the relaxation of the ionic atmosphere. 
Our theory also predicts that the dispersion in $\eta_{\rm eff}$ takes place in lower (normalized) frequencies $\Omega=\omega/D_{\rm R}\kappa^2$ as the asymmetry of the salt $|\gamma_{\rm R}|$ increases.

In Sec.~\ref{sec:conduc}, we have studied the frequency dependence of the electrical conductivity for not very large frequency such that $|\omega| \ll \Lambda \kappa^2$. At the one-loop level, there appears two contributions, the electrophoretic term $\sigma^{\rm ph}$ and the relaxation term $\sigma^{\rm rel}$. At zero-frequency, the sum of these reduces to the classical result of Onsager\cite{Onsager1927,RobinsonStokes}. For non-zero frequencies, while $\sigma^{\rm rel}$ has been derived many years ago by Debye and Falkenhagen\cite{DebyeFalken}, the explicit expression of the electrophoretic term $\sigma^{\rm ph}$ has been missing in the literature. For low frequencies, $\omega\ll \nu_0\kappa^2$, the frequency dependence is dominated by $\sigma^{\rm rel}$ and thus the real part $\sigma'$ increases with increasing $\omega$ (Debye-Falkenhagen effect). As the frequency is further increased, electrophoretic term $\sigma^{\rm ph}$ becomes relevant, and the real part $\sigma'$ exhibits a maximum at the wavenumber $\omega_{\rm max}$ as a result of the competition of electrophoretic and relaxation terms. Our theory predicts that the maximum (angular) frequency $\omega_{\rm max}$ is proportional to $\kappa^2\propto \bar n$ as a \textit{universal feature} in dilute electrolyte solutions. For reasonable parameter values, the proportionality coefficient is about $10^2D_{\rm R}$, i.e.,  $\omega_{\rm max}\approx 10^2 D_{\rm R}\kappa^2$. 

Finally, in Sec.~\ref{sec:discussion}, we have discussed how higher-order terms in the loop-expansion and the inter-ionic short-range interactions contribute, as an example, to the zero-frequency effective viscosity $\eta_{\rm eff}(0)$. We have shown by a simple argument that the two-loop graph in Fig.~\ref{fig:graph_vis2}a contributes to the Jones-Dole $B$-coefficient, while the inter-ionic short-range interactions give no contribution to the $B$-coefficient at the one-loop level.

We make final remarks. More concentrated solutions, which are practically more relevant, may be studied in a similar manner by taking into account the inter-ionic short-range interactions\cite{Okamoto2020, Okamoto2021} as well as the higher-order loops. One can also extend the present theory to more complex mixture-solvent systems, taking into account the fluctuations of the solvent composition\cite{Okamoto2010,Onuki2011,Onuki2016}.  In this paper, most of the results are within the one-loop approximation. In the zero-frequency limit, the one-loop results for effective viscosity and conductivity reduce to the classical limiting laws that have also been (at least partially) derived from the linearized FHE\cite{Wada2005,Peraud2017,Donev2019}. Hence, we may expect that the one-loop results of the frequency dependence could also have been derived from the linearized FHE, without using the MSRJD formalism.
However, there are some advantages of using the MSRJD formalism:\\
(i) As has been mentioned in Sec.~\ref{sec:intro}, it naturally leads to the effective kinetic coefficients, i.e., $L_{ij}^{\rm eff}$ and $\eta_{\rm eff}$ for the dynamics of average variables.\\
(ii) It enables us to see more clearly the approximation(s) involved in the calculations. To be more concrete, we have seen that the one-loop approximation and the free energy density of the form in Eq.~(\ref{dilute}) are at least formally distinct approximations, though they are coherent with each other for dilute solutions. \\
(iii) It provides systematic means to improve the approximations, e.g., higher-order loops (though the perturbation calculations of higher orders would be extremely tedious). \\
(iv) Even without performing explicit calculations, as we have seen in Sec.~\ref{sec:discussion}, one can easily have some information of how some improvements of the approximations may contribute to the transport properties, e.g., the effective viscosity. \\
Besides the higher-order loops, the short-range ion-ion interaction becomes also relevant for concentrated solutions. As we have seen in Sec.~\ref{sec:discussion}, the short-range interactions yield the excess viscosity of order $\bar n^{3/2}$. It is also worth noting that the recent paper by Avni \textit{et al.} has shown that the DC conductivity derived from the linearized FHE (equivalently, the one-loop approximation) with the hard-sphere-like repulsion well fit the experimental data for concentrated solutions\cite{Avni2022}. Finally, we mention that the present path-integral approach, like many perturbative approach in quantum field theory, has no rigorous mathematical foundation though the same approach has extensively been used in the physics literature. For non-ionic fluctuating hydrodynamics, Donev \textit{et al.} have studied the nonlinear effect  on diffusion in a mathematically more controlled manner \cite{Donev2014}.
We wish the present study would stimulate future studies on the dynamics in more concentrated electrolyte solutions as well as mixture systems.

\begin{acknowledgments}
   The author would like to thank Hirofumi Wada and Tsuyoshi Yamaguchi for informative discussions. This work was supported by JSPS KAKENHI (grant nos. 18K03562 and 18KK0151).
\end{acknowledgments}
\vspace{3mm}



\appendix
\section{Drift term under It\^o prescription\label{sec:drift}}
In general, one must add a drift term to Eq.~(\ref{diffusion}) to ensure that the equilibrium probability distribution of the density fluctuations obeys the Boltzmann weight $\propto\exp(-{\cal F}/T)$. Under the It\^o prescription and with the choice of $L_i$ in Eq.~(\ref{kinetic_D}), the drift term is\cite{OnukiBook,Lau2007}
\begin{align}
   D_i \int_{{\bm r}'}\frac{\p{\cal M}_i({\bm r},{\bm r}')}{\delta n_i({\bm r}')} \label{drift}
\end{align} 
with ${\cal M}_i({\bm r},{\bm r}')=\nabla'\cdot\nabla n_i({\bm r})\delta_{{\bm r}-{\bm r}'}$.
In terms of ${\cal M}$, Eqs.~(\ref{diffusion}) and (\ref{noise_corr}) are respectively written as
\begin{align}
    &\frac{\p n_i}{\p t}=-{\bm v}\cdot\nabla n_i-D_i\int_{{\bm r}'}{\cal M}({\bm r},{\bm r}')\frac{\delta ({\cal F}/T)}{\delta n_i({\bm r}')}+\eta_i \label{diffusion_app}\\
    &\av{\eta_i(\underline r)\eta_j(\underline r')}=2\delta_{ij}D_i{\cal M}({\bm r},{\bm r}')\delta_{t-t'}.
\end{align}
We perform the integral in Eq.~(\ref{drift}) to show the drift term vanishes:
\begin{align}
    \int_{{\bm r}'}\frac{\p{\cal M}_i({\bm r},{\bm r}')}{\delta n_i({\bm r}')} 
    =&\nabla\cdot \int_{{\bm r}'}\nabla' \frac{\delta n({\bm r})}{\delta n({\bm r}')} \delta_{{\bm r}-{\bm r}'} \nn\\
    =&2\nabla\cdot \int_{{\bm r}'}\delta_{{\bm r}-{\bm r}'}\nabla' \delta_{{\bm r}-{\bm r}'}\nn\\
    =&-2\nabla\cdot \int_{{\bm r}'}\delta_{{\bm r}-{\bm r}'}\nabla' \delta_{{\bm r}-{\bm r}'}\nn\\
    =&0.
\end{align}
In a mathematically strict sense, the calculation in the above cannot be justified because the products of delta functions are mathematically ill-defined, and hence equations like Eq.~(\ref{diffusion_app}) need to be properly discretize.
It is worth noting that Kim \textit{et al}. have advocated a space-time discretization scheme of reaction-diffusion stochastic partial differential equations under It\^o prescription \cite{Kim2017}. For stochastic differential equations of finite dimensions with multiplicative noises, H\"utter and \"Ottinger have proposed the ``kinetic'' stochastic integral that leads to an efficient numerical scheme ensuring the correct equilibrium distribution \cite{Hutter1998}.

\section{Relation between effective coefficients and vertex functions\label{sec:ver_eff}}
In the present scheme, the effective coefficients are extracted from the fluctuations in equilibrium rather than the average currents in weakly non-equilibrium situations. To simplify the argument, let us demonstrate how the vertex functions are related to the effective coefficients in the case of one dynamical variable $x(t)$ (More complete discussion for field variables can be found in Ref.\cite{Tauber}). Let $F(x)$ denote the free energy that may include non-Gaussian contributions. We can symbolically write the Langevin equation governing $x(t)$ as
\begin{align}
    \frac{dx}{dt}=-L(x)\frac{\p (F/T)}{\p x}+\eta(t,x) \label{example}
\end{align}
with $\av{\eta(t,x(t))\eta(t',x(t'))}=2L(x(t))\delta(t-t')$.
As in the same manner as in the main text, one can introduce the auxiliary variable $\tilde x(t)$, the correlation functions $2\pi \delta_{\omega+\omega'}C(\omega)=\av{x(\omega)x(\omega')}$ and $2\pi \delta_{\omega+\omega'}G(\omega)=\av{x(\omega)\tilde x(\omega')}$, and the corresponding vertex functions $\Gamma^{(1,1)}(\omega)$ and $\Gamma^{(0,2)}(\omega)$. We define the following matrices,
\begin{align}
    {\cal G}(\omega)\equiv \begin{pmatrix}
      C(\omega) & G(\omega)\\
      G(-\omega) & 0
    \end{pmatrix},\quad
    \Gamma(\omega)\equiv \begin{pmatrix}
        0& \Gamma^{(1,1)}(\omega)\\
        \Gamma^{(1,1)}(-\omega) & \Gamma^{(0,2)}(\omega)
    \end{pmatrix}.
\end{align}
The key to the effective coefficients is the following formal relation:
\begin{align}
    {\cal G}(\omega) \Gamma(\omega)=1,
\end{align}
or, more explicitly,
\begin{align}
    G(\omega)=\Gamma^{(1,1)}(\omega)^{-1}, \quad C(\omega)=-\Gamma^{(1,1)}(-\omega)^{-1}\Gamma^{(0,2)}(\omega)\Gamma^{(1,1)}(\omega)^{-1},  \label{identity}
\end{align}
which can be interpreted as generalizations of Eqs.~(\ref{G_p0})--(\ref{C_v0}) for the bare propagators to the full correlation functions.
The derivation can be found in Refs.~\cite{Tauber,Amit} or any textbook on the quantum field theory.
The same relation holds for more general multi-variable/field-variable cases with $\Gamma^{(1,1)}(\omega)^{-1}$ in Eq.~(\ref{identity}) being the inverse matrix/operator. Note also that it is not based on perturbation theories, so it holds up to any order in the loop expansion.

Meanwhile, we may write the effective linear Langevin equation for $x$ that is consistent with the two-point correlation functions $G$ and $C$:
\begin{align}
    \frac{d x}{dt}= -L_{\rm eff}\frac{\p (F_{\rm eff}/T)}{\p x}+\eta_{\rm eff}(t) \label{example_eff}
\end{align}
with $\av{\eta_{\rm eff}(t)\eta_{\rm eff}(t')}=2L_{\rm eff}\delta(t-t')$. In the above, $L_{\rm eff}$ and $F_{\rm eff}=(T\mu_{\rm eff}/2)x^2$ are the effective kinetic coefficient and the effective (Gaussian) free energy, respectively, which include the effects of non-Gaussian terms in $F$ and the multiplicative noise in Eq.~(\ref{example}). 
Here we have assumed the zero-frequency limit of $L_{\rm eff}$ but the generalization to finite frequencies is straightforward. According to the Onsager's regression hypothesis, Eq.~(\ref{example_eff}) without the noise term coincides with the macroscopic deterministic law. The equation (\ref{example_eff}) is a linear equation with an additive noise, so we can readily obtain the correlation functions,
\begin{align}
    G(\omega)=\frac{1}{i\omega+L_{\rm eff}\mu_{\rm eff}},\quad C(\omega)=\frac{2L_{\rm eff}}{\omega^2+(L_{\rm eff}\mu_{\rm eff})^2}. \label{onsg}
\end{align}
Now, comparing Eq.~(\ref{identity}) with (\ref{onsg}), we have
\begin{align}
    \Gamma^{(1,1)}=i\omega+L_{\rm eff}\mu_{\rm eff}, \quad \Gamma^{(0,2)}=-2L_{\rm eff}. \label{example_gamma}
\end{align}
That is, the vertex function $\Gamma^{(0,2)}$ yields the effective kinetic coefficient $L_{\rm eff}$, while Eq.~(\ref{example_gamma}) also leads to the effective free energy, $-2\Gamma^{(1,1)}(0)/\Gamma^{(0,2)}=\mu_{\rm eff}$. 

\section{Calculation of $\Delta{\cal L}_{ij}^{\rm a}$ \label{sec:Lija}}
To calculate $\Delta{\cal L}_{ij}^{\rm a}$, we need only the terms proportional to $q^2$. In the left graph of Fig.~\ref{fig:L}a, the two vertices have the factors, $v_{k;i}({\bm q}_1,-{\bm q}-{\bm q}_1)$ and $v_{l;j}(-{\bm q}_1,{\bm q}+{\bm q}_1)$ which are defined in Eqs.~(\ref{v_ij1})--(\ref{v_ij4}). Each of these factors is at least of order $q$, so we can neglect the terms of order $q^2$ in each factor that are irrelevant to $\Delta {\cal L}_{ij}$. For $k,l=1$, we have to the first order in $q$,
\begin{align}
    &v_{1;i}({\bm q}_1,-{\bm q}-{\bm q}_1)=\left\{ \begin{array}{lr}
        4\pi\ell_{\rm B}\gamma D q_1^{-2} ({\bm q}_1\cdot {\bm q}) & (i=1) \\
        4\pi\ell_{\rm B} D q_1^{-2} ({\bm q}_1\cdot {\bm q}) & (i=2)
    \end{array}\right. \label{v_1i}\\
    &v_{1;j}(-{\bm q}_1,{\bm q}+{\bm q}_1)=\left\{ \begin{array}{lr}
        4\pi\ell_{\rm B}\gamma D q_1^{-2} ({\bm q}_1\cdot {\bm q}) & (j=1) \\
        4\pi\ell_{\rm B} D q_1^{-2} ({\bm q}_1\cdot {\bm q}) & (j=2)
    \end{array}\right. .
\end{align}
 Similarly in the right graph of Fig.~\ref{fig:L}b, the two vertices have the factors, $v_{k;i}({\bm q}_1,-{\bm q}-{\bm q}_1)$ and $v_{l;j}({\bm q}+{\bm q}_1,-{\bm q}_1)$. The former factor for $k=1$ is the same as Eq.~(\ref{v_1i}), while the latter is
\begin{align}
    v_{1;j}({\bm q}+{\bm q}_1,-{\bm q}_1)=\left\{ \begin{array}{lr}
        -4\pi\ell_{\rm B}\gamma D q_1^{-2} ({\bm q}_1\cdot {\bm q}) & (j=1) \\
        -4\pi\ell_{\rm B} D q_1^{-2} ({\bm q}_1\cdot {\bm q}) & (j=2).
    \end{array}\right. \label{v_1j}
\end{align}
For $k,l=2$, these factors are of order $q^2$, so we only need to calculate the graphs for $k=l=1$. Hence for $i=j=1$ we obtain Eq.~(\ref{L11a0}). From Eqs.~(\ref{v_1i})--(\ref{v_1j}) we readily find $\Delta{\cal L}_{22}^{\rm a}=\Delta{\cal L}_{11}^{\rm a}/\gamma^2$ and $\Delta{\cal L}_{12}^{\rm a}=\Delta{\cal L}_{21}^{\rm a}=\Delta{\cal L}_{11}^{\rm a}/\gamma$, and thus have Eqs.~(\ref{L22a}) and (\ref{L12a}).

\section{Calculation of $\Delta{\cal L}_{ij}^{\rm b}$ \label{sec:Lijb}}
The left graph in Figs.~\ref{fig:L}b has the two vertex factors $v_{k;i}({\bm q}_1,-{\bm q}-{\bm q}_1)$ and $u_{m;j,l}({\bm q},{\bm q}_1)$. We have already discussed the former in Appendix \ref{sec:Lija}. The latter factor is obtained from Eqs.~(\ref{vertex_u1}) and (\ref{vertex_u2}) as
\begin{align}
    &u_{1;11}({\bm q},{\bm q}_1)=u_{2;12}({\bm q},{\bm q}_1)=u_{1;22}({\bm q},{\bm q}_1)=2D{\bm q}\cdot{\bm q}_1 \label{vertex_u1_2}\\
    &u_{2;11}({\bm q},{\bm q}_1)=u_{1;12}({\bm q},{\bm q}_1)=u_{2;22}({\bm q},{\bm q}_1)=2D\gamma {\bm q}\cdot{\bm q}_1,\label{vertex_u2_2}
\end{align}
which are all of order $q$. Hence we need to calculate the graphs only for $k=1$ because the graphs for $k=2$ are of order $q^3$. Similarly, the right graph in Figs.~\ref{fig:L}b have two vertex factors $v_{k;i}(-{\bm q}-{\bm q}_1,{\bm q}_1)$ and $u_{m;j,l}({\bm q},{\bm q}_1)$, where the latter is already given in Eqs.~(\ref{vertex_u1_2}) and (\ref{vertex_u2_2}), and the former is given by
\begin{align}
    v_{1;i}(-{\bm q}-{\bm q}_1,{\bm q}_1)=\left\{ \begin{array}{lr}
        -4\pi\ell_{\rm B}\gamma D q_1^{-2} ({\bm q}_1\cdot {\bm q}) & (i=1) \\
        -4\pi\ell_{\rm B} D q_1^{-2} ({\bm q}_1\cdot {\bm q}) & (i=2).
    \end{array}\right. \label{v1i_2}
\end{align}
One should note that for each $(i,j)$ there are also two more graphs that are the ones in Fig.~\ref{fig:L}b with the indices $i$ and $j$ being exchanged. For $i=j$ those flipped graphs clearly have the same contributions.
For $i=j=1$, we hence have
\begin{align}
    \Delta{\cal L}_{11}^{\rm b}=8\pi \gamma \ell_{\rm B} D^2 \int_{\ulq_1} (\hat{\bm q}\cdot \hat{\bm q}_1)^2\big[&\gamma C_{\phi 0}^{22}G_{\phi 0}^{11} +C_{\phi 0}^{22}G_{\phi 0}^{12} +C_{\phi 0}^{12}G_{\phi 0}^{11}+\gamma C_{\phi 0}^{12}G_{\phi 0}^{12} \nn\\
    &-\gamma C_{\phi 0}^{12}G_{\phi 0}^{21} -C_{\phi 0}^{12}G_{\phi 0}^{22} -C_{\phi 0}^{11}G_{\phi 0}^{21}-\gamma C_{\phi 0}^{11}G_{\phi 0}^{22}\big],
\end{align}
where the arguments of $G_{\phi 0}$'s and $C_{\phi 0}$'s are all $\ulq_1$. Performing the $\omega_1$-integration and using Eq.~(\ref{eigen}), we obtain Eq.~(\ref{L11b}). 
Similarly we have 
\begin{align}
    \Delta{\cal L}_{22}^{\rm b}=8\pi \gamma \ell_{\rm B} D^2 q^2 \int_{\ulq_1} (\hat{\bm q}\cdot \hat{\bm q}_1)^2\big[& C_{\phi 0}^{22}G_{\phi 0}^{11}+\gamma C_{\phi 0}^{22}G_{\phi 0}^{12} +\gamma C_{\phi 0}^{12}G_{\phi 0}^{11}+C_{\phi 0}^{12}G_{\phi 0}^{12} \nn\\
    &- C_{\phi 0}^{12}G_{\phi 0}^{21} 
    -\gamma C_{\phi 0}^{12}G_{\phi 0}^{22} -\gamma C_{\phi 0}^{11}G_{\phi 0}^{21}-C_{\phi 0}^{11}G_{\phi 0}^{22} \big],
\end{align}
which yields Eq.~(\ref{L22b}).

\section{Derivation of Eqs.~(\ref{eta_excess}) and (\ref{H_scaling}) \label{deriv_H}}
Using Eqs.~(\ref{Cp0}) and (\ref{eigen}), we perform the $\omega_1$-integration in Eq.~(\ref{self_v2}),
\begin{align}
    \int_{\omega_1} C_{\phi 0}^{22}({\bm q}_1,\omega_1)C_{\phi 0}^{22}({\bm q}_1,\omega_1+\omega) 
    = \frac{16\bar n^2D^2q_1^4}{(\lambda_1+\lambda_2)^2(\lambda_1-\lambda_2)^2} \Big[ &\frac{\lambda_1^3(\lambda_2^2 g-1)^2}{\omega^2+4\lambda_1^2}+\frac{\lambda_2^3(\lambda_1^2 g-1)^2}{\omega^2+4\lambda_2^2} \nn\\
    &\hspace{-15mm}-\frac{\lambda_1\lambda_2(\lambda_1+\lambda_2)(\lambda_2^2 g-1)(\lambda_1^2 g-1)}{\omega^2+(\lambda_1+\lambda_2)^2}\Big],\label{CC_omega}
\end{align}
where $\lambda_i=\lambda_i(q_1)$, and we have defined
\begin{align}
    g(q_1)\equiv \frac{1}{D^2(1-\gamma^2)(q_1^2+\kappa^2)^2}.
\end{align}
Using Eq.~(\ref{eigen}), we can rewrite this expression in terms of $\Omega=\omega/D\kappa^2$, $s$, and $\hat \lambda_i$ defined in Eqs.~(\ref{dless0}) and (\ref{dless}):
\begin{align}
    \int_{\omega_1} C_{\phi 0}^{22}({\bm q}_1,\omega_1)C_{\phi 0}^{22}({\bm q}_1,\omega_1+\omega)= &\frac{16\bar n^2  q_1^4}{D\kappa^6 (s^2+1)(4\hlamb_1^2+\Omega^2)(4\hlamb_2^2+\Omega^2)\{(2s^2+1)^2+\Omega^2\}} \nn\\
    &\times\big[-4(1-\gamma^2)s^4(2s^2+1)\{\gamma^2(s^2+1)-(2s^2+1)\} \nn\\
    &\hspace{6mm}+\Omega^2\{ (2s^2+1)(5s^2+1)-(1-\gamma^2)(2s^4+3s^2)\}\nn\\
    &\hspace{6mm}+\Omega^4
    \big].\label{CC_omega2}
\end{align}
Meanwhile, we also perform the angular-integration in Eq.~(\ref{self_v2}) as
\begin{align}
    &\int_0^{2\pi}d\phi\int _0^\pi d\theta\, \sin\theta (\hat{\bm q}\cdot\hat{\bm q}_1)^2[1-(\hat{\bm q}\cdot\hat{\bm q}_1)^2] =2\pi \int _0^\pi d\theta\, \cos^2\theta\, \sin^3\theta=\frac{8\pi}{15}. \label{int_angular}
\end{align}
Substitution of Eqs.~(\ref{CC_omega2}) and (\ref{int_angular}) into (\ref{self_v2}) yields Eqs.~(\ref{eta_excess}) and (\ref{H_scaling}).

\section{Calculation of the maximum frequency\label{sec:omega_max}}
We here calculate the reduced maximum frequency $\Omega_{\rm max}$ as a function of $\hat\nu_0$ and $C_\sigma$. The explicit expression for $g'_\sigma$ can be obtained from Eq.~(\ref{g_def}) as
\begin{align}
    g'_\sigma=\frac{-C_\sigma}{(1+\Omega^2/\hat\nu_0^2)} \Big[ u_g^{3/4} \big\{\frac{\Omega}{\hat\nu_0}\sin v_g -\cos v_g\big\} +1\Big] +(\mbox{relaxation term}),
\end{align}
where the ``(relaxation term)'' is the real part of the second term of Eq.~(\ref{g_def}), and we have defined $u_g=(\Omega^2+1)/\hat\nu_0^2$ and $v_g=(3/2)\arctan\Omega$.
Anticipating the maximum appears at a high frequency, $\Omega_{\rm max} \gg 1$, we make the approximations, $u_g\approx \Omega^2/\hat\nu_0^2$, $v_g\approx 3\pi {\rm sgn}(\Omega)/4$ and (relaxation term)$\approx -|\Omega|^{-1/2}$. Then, in terms of the new variable $x=|\Omega/\hat\nu_0|^{1/2}$, we can write $g'_\sigma$ as
\begin{align}
    \hat\nu_0^{1/2}g_\sigma'\approx -\frac{\epsilon_\sigma^{-1}}{1+x^4}\Big[ x^3 (x^2+1)+\sqrt{2} \Big] -x^{-1}
\end{align}
where $\epsilon_\sigma=\sqrt{2}/(\hat\nu_0^{1/2}C_\sigma)$ is the small parameter defined below Eq.~(\ref{Omega_max_app}). The extremum condition $\p g'_\sigma/\p x=0$ yields
\begin{align}
    0=(-x^{6}+x^4-5x^2+4\sqrt{2}x-3)x^4+\epsilon_\sigma(x^8+2x^4+1). \label{max_eq}
\end{align}
We seek the approximate solution for small $\epsilon_\sigma$. At $\epsilon_\sigma=0$, the above equation has a trivial root $x=0$. For finite $\epsilon_\sigma$ we find the expansion of the form, $x=\epsilon_\sigma^{1/4} x_{1/4}+\epsilon_\sigma^{1/2} x_{1/2}\cdots$. Substituting this form into Eq.~(\ref{max_eq}), we obtain $x_{1/4}=3^{-1/4}$ from the terms proportional to $\epsilon_\sigma$.
Similarly, the terms proportional to $\epsilon_\sigma^{5/4}$ yield $x_{1/2}=\sqrt{6}/9$, and hence
\begin{align}
    |\Omega_{\rm max}/\hat\nu_0|^{1/2}=\Big(\frac{\epsilon_\sigma}{3}\Big)^{1/4}+\frac{\sqrt{6\epsilon_\sigma}}{9}+\cdots.
\end{align}
This readily yields Eq.~(\ref{Omega_max_app}).
\bibliography{ref}
\end{document}